# Si–incorporated amorphous indium oxide thin–film transistors

Shinya Aikawa[1]*, Toshihide Nabatame[2], and Kazuhito Tsukagoshi[2]*

[1]*Department of Electrical and Electronic Engineering, Kogakuin University, Hachioji, Tokyo 192-0015, Japan*

[2]*International Center for Materials Nanoarchitectonics (WPI-MANA), National Institute for Materials Science (NIMS), Tsukuba, Ibaraki 305-0044, Japan*

E-mail: aikawa@cc.kogakuin.ac.jp, TSUKAGOSHI.Kazuhito@nims.go.jp

**Amorphous oxide semiconductors, especially indium oxide-based (InO$_x$) thin-films, have been major candidates for high mobility with easy-to-use device processability. As one of the dopants in InO$_x$ semiconductors, we proposed Si to design a thin-film transistor (TFT) channel. Because the suppression of unstable oxygen vacancies in InO$_x$ is crucial to maintaining the semiconducting behavior, Si was selected as a strong oxygen binder that is reasonably available for large production. In this review, we focus on the overall properties observed in Si-incorporated amorphous InO$_x$ TFTs in terms of bond-dissociation energy, Gibbs free energy, Si-concentration dependence of TFT properties, carrier transport mechanism, and bias stress instability. In comparing low and high doping densities, we found that the activation energy and density of states decreased at a high Si concentration in InO$_x$ TFTs, implying that the trap density was reduced. As a result, stable operation under bias stresses could be realized. Furthermore, the inverse Meyer-Neldel rule was observed in the highly Si-doped InO$_x$ TFT, indicating reasonable ohmic contact. Based on our fundamental knowledge of the**

**Si-doped InO$_x$ film, we developed a high-mobility bilayer TFT with a homogeneous stacked channel that was different from a TFT with an etch stop layer structure. The TFT showed remarkably stable operation. With simple element components based on InO$_x$, it is possible to systematically discuss vacancy engineering in terms of conduction properties.**

## 1. Introduction

High-mobility thin-film transistors (TFTs) are in high demand for high-speed switching operations in next-generation large-area flat panel displays (FPDs). The typical field-effect mobility of hydrogenated amorphous silicon (*a*-Si:H), which is extensively used for TFT backplanes in liquid crystal displays, is limited to < 1 cm$^2$/Vs.[1] Electron transport is strongly perturbed in the conduction band of Si, where carrier transport paths consist of $sp^3$ orbitals with strong directivity. The resulting bond angle fluctuation significantly alters the electronic levels, leading to high density trap states.[2] On the other hand, although the fabrication process is compatible with conventional amorphous-Si TFTs, amorphous indium oxide (InO$_x$) has far superior electrical properties compared with *a*-Si:H owing to its unique carrier transport properties.[3, 4] The conduction bands in InO$_x$-based materials are known to originate from the 5*s* orbital of In atoms, and the spherically spreading orbital suppresses the perturbing effect.[5-7] It is widely believed that InO$_x$-based thin-films have advantages in mobility over *a*-Si:H films. Furthermore, since metal oxides exhibit deformability of the amorphous phase and low temperature processability, flexible and transparent TFTs with reliable operation can be fabricated on plastic substrates if metal oxide materials are used for the active and dielectric layers and the electrodes.[2, 8, 9] Thus, InO$_x$-based TFTs are expected to be the most promising candidate for next-generation switching TFTs.

There have been many reports on $InO_x$-based TFTs such as In-O,[10-12] In-Al-O,[13] In-Ga-O,[14, 15] In-Ge-O,[16] In-Hf-O,[17] In-Si-O (ISO),[18, 19] In-Sn-O (ITO),[20, 21] In-W-O (IWO),[22, 23] In-Zn-O (IZO),[24, 25] multicomponent In-B-Zn-O,[26] In-C-Si-O,[27] In-C-W-O,[28] In-C-Zn-O,[29] In-Ga-Zn-O (IGZO),[2, 8] In-Hf-Zn-O,[30] In-Sc-Zn-O,[31] In-Si-W-O,[32] In-Si-Zn-O,[33-35] In-Sn-Zn-O,[9, 36] In-Ta-Zn-O,[37, 38] In-W-Zn-O,[39] and In-Zr-Zn-O[40]. Parthiban and Kwon have reported a discussion regarding the role of dopant in $InO_x$ in various elements so far.[41]

IGZO has received particular attention and has been used in large-size high-definition flat-panel-displays.[42, 43] However, there remain certain issues for improvement toward mass-production.[3, 5, 44] One issue is the formation of oxygen deficiencies, which causes instability in the electronic properties. Gallium in IGZO is known to suppress the formation of oxygen vacancies ($V_O$), and thus, addition of Ga is an effective method to overcome instability.[45-47] Components of amphoteric $Ga_2O_3$ and/or ZnO containing IGZO are sensitive to wet etching processes. Thus, an alternative to develop Ga- and Zn-free amorphous oxide transistors is required. Displacement of Ga and/or Zn to other elements is an effective approach for the suppression of $V_O$ in an $InO_x$ film. In addition, IGZO TFTs are susceptible to environmental factors such as oxygen and moisture and have problems with long-term stability.[3, 48, 49] When oxygen molecules are desorbed from an oxide semiconductor because of changes in the atmospheric environment, the TFT properties are affected.[50-52] This is an issue for processes that occur in a reducing atmosphere[53] such as $SiN_x$ passivation[54] and dry etching of the source/drain electrodes.[55]

Recently, we proposed that W- and/or Si-doped $InO_x$ is suitable for high-mobility TFTs with preferable stable electrical behavior.[18, 19, 22, 23, 32, 56, 57] The homogeneous bilayer TFT developed based on the results showed field effect mobility of 19.6 cm$^2$/Vs, and the shifts of turn-on voltages ($V_{ON}$) under positive and negative bias stress instability measurements were 0.2 and 0.8 V, respectively. The successful demonstration above of high mobility and stability is due to the control of the $V_O$ formation. An appropriate oxygen binding dopant is required to obtain such reliability. Here, we report our concept of dopant selection in $InO_x$ semiconductors from the viewpoint of bond-dissociation energy. Then, electrical properties and electron transport in the Si-incorporated amorphous $InO_x$ TFTs based on Si-concentration dependence and temperature dependent $I$-$V$ characterization are discussed. Finally, the aforementioned bilayer TFT is introduced in high-performance operation.

## 2. Device fabrication and characterization

Our TFTs for electric characterization were fabricated by the following typical procedure. The $InO_x$-based TFTs were fabricated on a heavily doped p-type Si substrate with a thermal $SiO_2$ layer. The substrate was first cleaned using a process of ultrasonication in acetone and isopropyl alcohol, followed by UV-ozone exposure. An $InO_x$-based semiconductor film with a thickness of 10 nm was then deposited by DC magnetron sputtering at room temperature with a various $O_2$/(Ar + $O_2$) ratios. The sputtering power and working pressure were optimized depending on the sputtering target compositions. We emphasize that all sputtering targets used in our study were completely Zn-free. The channel width ($W$ = 1000 μm) was defined by using a stencil shadow mask. After deposition, the post annealing process was performed. Source and drain electrodes (40 nm) were formed by vacuum evaporation (thermal or electron beam) through a stencil shadow mask, which defined the channel lengths ($L$) from 50 to 350 μm with an interval of 50 μm. The devices had a simple bottom-gate top-contact TFT configuration without any passivation layer.

The channel thickness was confirmed with an ellipsometer, and the film morphology was observed by atomic force microscopy (AFM) in tapping mode. The electrical characteristics of the fabricated TFTs were measured in the dark at room temperature using a precision semiconductor parameter analyzer (Agilent 4156C) connected to a probe station placed in a high vacuum chamber. After the characterization, a series of TFT parameters, i.e., field-effect mobility in the linear regime ($\mu_{FE}$), saturation field-effect mobility ($\mu_{sat}$), subthreshold swing ($ss$), and $V_{ON}$, on- and off-state current ratio ($I_{ON}/I_{OFF}$) were obtained. The $\mu_{FE}$, $\mu_{sat}$ and $ss$ were extracted using the following equations:

$$\mu_{FE} = \frac{\partial I_D}{\partial V_{GS}} \frac{L}{W} \frac{1}{C_i V_{DS}}, \qquad (1)$$

$$\mu_{sat} = \left(\frac{\partial \sqrt{I_D}}{\partial V_{GS}}\right)^2 \frac{L}{W} \frac{2}{C_i}, \qquad (2)$$

$$ss = \left(\frac{\partial \log_{10} I_D}{\partial V_{GS}}\right)^{-1}, \qquad (3)$$

where $L$ and $W$ are the channel length and width, respectively, and $C_i$ is the gate capacitance per unit area.

# 3. Strategies for choosing dopants in InO$_x$ semiconductor

## 3.1 W-doping using acid insoluble WO$_3$

We first focus on W-doped InO$_x$ (IWO), since the constituent WO$_3$ is insoluble in acids except for hydrogen fluoride solutions.[58] IWO is known as transparent conductive oxide films.[59-62] Although there have been several studies on IWO electrodes for organic light-emitting diodes,[63] organic solar cells,[64] and flexible carbon-nanotube transistors,[65] reports focusing on the significant semiconducting properties of this material are limited.[66-69]

Figure 1(a) shows a schematic TFT structure. The inset is a roughness profile of the IWO film observed by AFM after 3 times annealing at 100 °C for 5 min in N$_2$ using rapid thermal anneal equipment. The use of $N_2$ annealing is due to the improvement of TFT properties as has been reported for IGZO.[70, 71] In this case, the $In_2O_3/WO_3$ ratio in the IWO target was 99/1 wt. %. The root mean square (RMS) roughness measured was 0.27 nm, indicating that the surface was almost as flat as the Si substrate (0.24 nm). X-ray diffraction (XRD) and cross-sectional transmission electron microscopy (TEM) together with selected area electron diffraction indicate that the IWO film remained amorphous after the annealing treatment. The ionic radii of $W^{6+}$ and $In^{3+}$ are 0.06 and 0.08 nm, respectively, for the same coordination number.[72] This difference is known to induce a distortion in $In_2O_3$ crystals; thus, the structural role of W atoms in $InO_x$ is of amorphization, leading to a very flat and smooth surface. Figure 1(b) and (c) show typical output ($I_D$-$V_{DS}$) and transfer ($I_D$-$V_{GS}$) characteristics. A clear pinch-off behavior, small hysteresis (~0.7 V) and low $I_{OFF}$ (~$10^{-14}$ A) can be seen. At $V_{DS}$ = 40 V, the estimated $\mu_{sat}$, $I_{ON}/I_{OFF}$ and $ss$ are 19.3 cm²/Vs, 8.9×10⁹ and 0.47 V/decade, respectively. The $\mu_{sat}$ value was calculated using Eq. (2). The $C_i$ in this case is estimated to be 1.73×$10^{-8}$ F/cm² based on a dielectric constant of 3.9 for $SiO_2$.

## 3.2 Bond-dissociation energy with oxygen

A target of the IWO TFT development was to realize a stable TFT without a passivation layer by using acid resistant oxides. Fortunately, W-O binding has strong bond-dissociation energy (BDE) and is more rigid to acid. For further effective BDE element, we use three elements, W, Ti, or Si, that have high BDE and compare $InO_x$-based films doped with these elements. We demonstrate that considering the BDE is useful when selecting appropriate dopants for $InO_x$-based semiconductors.

The Ti-, W-, and Si-doped $InO_x$ sputtering targets contained 1 wt. % $TiO_2$, $WO_3$, and $SiO_2$, respectively, hereafter denoted ITiO, IWO, and ISO, respectively. After the sputtering deposition of these channel layers at various oxygen partial pressures ($P_{O2}$), an Au source/drain electrode was formed.[18] Figure 2 shows the electrical conductivity of the as-deposited films as a function of the $P_{O2}$ used during sputtering deposition. Each value was extracted from the linear region of the output characteristics when carriers are highly induced ($V_{GS}$ = 30 V). Similar to other oxide semiconductors such as IGZO[73, 74] and ZnO,[75] the electrical conductivity of the InO$_x$-based films depends on the $P_{O2}$; this behavior is caused by the strong dependence of the carrier concentration during deposition.[8, 76] Although the dopant concentration is only 1 wt. %, the choice of dopant species strongly affects the electrical conductivity. For the ITiO, the electrical conductivity changes significantly based on $P_{O2}$; however, this behavior does not occur in the ISO film. The effect of Si is greater than that of Ti because the atomic ratios of Si/In in ISO and Ti/In in ITiO are 2.3 and 1.8 %, respectively. On the other hand, the atomic ratio of W/In in IWO is 0.6 %, and even though ratio is much smaller than that of Ti, the change in conductivity is rather small. The ISO exhibits the most stable electrical conductivity among the three films, indicating that the slope of the electrical conductivity as a function of the $P_{O2}$ could be related to the BDE of the dopant. The BDE is defined as the strength of the chemical bond determined when a diatomic species decomposes into individual atoms. If the energy of the metal-to-oxygen bond (X–O, where X is the dopant material) is low, then oxygen can be easily released, which increases the carrier density.[8, 42] In contrast, if a dopant with a high BDE is introduced to InO$_x$, we expect the electrical properties of the oxide film to be more stable over the range of $P_{O2}$ used during deposition (the inset of Fig. 2). The conductivity slopes become flatter in the order of ITiO, IWO, and ISO, which corresponds to the order of their BDE: 666.5, 720, and 799.6 kJ/mol for Ti-O,[77] W-O,[78] and Si-O,[79] respectively. These changes in the conductivity strongly support that the BDE can be used to determine the role of the dopant in $InO_x$-based semiconductors.

Figures 3(a)–(f) show the typical electrical properties of the fabricated TFTs annealed at 150 °C in air. Figures 3(a), (c), and (e) show the output characteristics ($I_D$–$V_{DS}$) for the three TFTs, each with a different dopant. The transfer characteristics ($I_D$–$V_{GS}$) in the saturation region ($V_{DS}$ = 30 V) for ITiO, IWO, and ISO have negligibly small hystereses and high $I_{ON}/I_{OFF}$. The transistor properties observed in three TFTs are summarized in Table 1. The $\mu_{sat}$ are 32, 30, and 17 $cm^2$/Vs for the TFTs fabricated using the ITiO, IWO, and ISO targets, respectively. The field-effect mobility for ISO is slightly suppressed, which may be due to the incorporation of the strong binder Si into $InO_x$. However, the comparison of precisely controlled W and Si dopant concentrations shows that the mobility of ISO is higher than that of IWO[57] because Si has a smaller ionic radius than W regardless of coordination number or valence state,[80] suggesting that the scattering cross section of Si is smaller than that of W.[31] In addition, the most significant scattering in $InO_x$ systems is ionic scattering.[81] A charge carrier screening effect in which the Coulomb potential of the ion core hole creates localized trap states by pulling an orbital out of the conduction band[82, 83] is reduced in an ionic scattering source.[84] Thus, the higher mobility of the ISO may have been obtained because a higher carrier density is induced in the ISO compared with the IWO at the same $V_{GS}$. Parthiban and Kwon reported that dopants with high Lewis acid strength effectively enhance charge carrier screening.[26] Such dopants polarize electronic charges away from the $2p$ valence band, where the polarity is dependent on the Lewis acid strength. Because $Si^{4+}$ has a higher Lewis acid strength (8.096) than $W^{6+}$ (3.158), the higher mobility realized by the charge carrier screening is confirmed to be compatible with the discussion regarding the Lewis acid strength of the dopants.

By comparing the mobilities of IGZO and IZO, it is reasonable that the lower mobility of IGZO could be caused by the incorporation of Ga, as suggested by Kamiya *et al.*[7] In IGZO, Ga-incorporation effectively stabilizes the TFT electrical properties because the BDE of Ga–O is 374 kJ/mol, a value slightly higher than that of In–O.[85] Thus, a large amount of Ga is needed in IZO for stabilization of the electrical properties. On the other hand, we examined the improvement of electrical stability by using dopants with higher BDE. The ISO fabricated TFT had higher electrical stability after annealing, even though the dopant density was rather low (1 wt. %).[18] According to first-principle calculations,[86] the electrical stability of amorphous oxide semiconductors can be effectively controlled by incorporating Si, because the formation of $V_O$ can be suppressed by the strong binding of Si–O.

## 4. Si-concentration dependence in ISO TFT

### 4.1 Series of TFT parameters

We consider the doping effect of Si in $InO_x$ films. For systematic understanding, we prepared three sputtering targets with different Si concentration; 3, 5, and 10 wt. % $SiO_2$, denoted ISO-3, -5, and -10, respectively. The films were fabricated at common conditions: DC sputtering was performed at a sputtering power of 200 W under an $Ar/O_2$ atmosphere at 0.25 Pa with various $P_{O2}$. Then, the films were annealed at 250 °C for 30 min in air. After deposition of the Mo source and drain electrode, the devices were again annealed at 150 °C for 30 min in air and then at the same temperature for 5 min in $O_3$.

X-ray diffraction characterization showed that all the films were in the amorphous phase against thermal treatments up to 350 °C. The crystallization temperature of the films became higher as the $SiO_2$ content ratio in ISO increased. The ISO-10 films were not crystallized even after thermal annealing at 600 °C. Because no Si clusters were observed, Si atoms were uniformly dispersed in an $In_2O_3$ matrix.[87] Figures 4(a) and (b) are typical transfer characteristics of the ISO-3 and ISO-10 TFTs. As shown in Fig. 4(c), a series of TFT parameters ($\mu_{FE}$, $I_{OFF}$, $I_{ON}$, $ss$, and $V_{ON}$) can be obtained as a function of $SiO_2$ content at various $P_{O2}$ from the linear region in the transfer curve.

The energy levels of In atoms with dangling bonds, i.e., $V_O$, are formed near the conduction band edge, which acts as shallow donor state. The formation of $V_O$ is suppressed by the incorporation of Si, whose BDE is much higher than that of the other constituents in the film. The suppression of $V_O$ by adding Si was previously reported in InZnO- and ZnSnO-based systems;[35, 86, 88, 89] however, the materials showed no semiconducting behaviors for thermal treatment processes above 200 °C. In the present case, the 250 °C thermotolerant and Zn-free $InO_x$-based amorphous semiconductor films were realized by heavily incorporating $SiO_2$ into $In_2O_3$ films. The TFT properties of oxide semiconductors are not only affected by the composition ratio of the target[45, 46] but also by the mixture ratio of $Ar/O_2$ during sputtering.[90, 91] The $\mu_{FE}$, $I_{OFF}$, $I_{ON}$, and $ss$ decreased as the $SiO_2$ content in the sputtering target and $O_2$ ratio in the $Ar/O_2$ gas increased. On the other hand, the $V_{ON}$ shifted from negative to positive $V_{GS}$ as the $SiO_2$ and $O_2$ ratio increased. These characteristic behaviors were similar to those observed through a reduction in $V_O$ in the $InO_x$-based semiconductor film.[92] Then, as the $SiO_2$ content ratio increased, the distribution of $ss$ and $V_{ON}$ against the $O_2$ ratio during sputtering decreased. Because the Si-O bonds are stronger than In-O bonds, the ISO with high $SiO_2$ content allows the small amount oxygen during sputtering to compensate the oxygen-deficient sites. We also clarify that the W has the same role as $V_O$ suppressor, but the thermotolerance is much different because of the different BDE of W.[23]

## 4.2 Comparison of low and high Si concentrations

For a better understanding of the role of Si, two TFTs of ISO-3 and ISO-10 are compared. To adjust the $V_{ON}$ to be 0 V, the $P_{O2}$ was optimized; $O_2/(Ar + O_2) = 50$ % for ISO-3, and $O_2/(Ar + O_2) = 8.3$ % for ISO-10. The bias stress instabilities were measured against $V_{GS} = \pm 30$ V for up to 3 hours (Fig. 5). The lack of degradation in the *ss* values suggests that unstable defects were highly suppressed in the ISO film. Negative shifts in the $V_{ON}$, which are attributed to depopulation of donor-like traps, were observed in all the devices.[93] Although a positive shift in $V_{ON}$ originating from charge trapping in the gate insulator under continuous positive bias stress (PBS) has been commonly observed in oxide semiconductor systems,[94, 95] no positive shift was obtained for the ISO, indicating that the incorporation of $SiO_2$ into $In_2O_3$ effectively reduces charge trapping.

The negative shift observed in $V_{ON}$ under PBS is considered to be triggered by Joule heating and the hot carrier effect.[96, 97] Application of higher $V_{GS}$ for a longer time resulted in a larger shift in the $V_{ON}$. The shift of $V_{ON}$ is related to the trap densities at the semiconductor/insulator surface ($N_S$) and the bulk of the film.[93] The $N_S$ can be approximated from the *ss* value using the following equation:[98]

$$N_s = C_i \left\{ \frac{(ss) \times \log_{10} e}{k_B T} - \frac{1}{q} \right\}, \qquad (4)$$

where $e$, $k_B$, $T$ and $q$ are the Napier constant, Boltzmann constant, absolute temperature, and elementary charge, respectively. For the ISO-3 and ISO-10 at 20 °C, $N_S$ of $2.50 \times 10^{11}$ /cm$^2$ and $1.08 \times 10^{11}$ /cm$^2$ are obtained, respectively. The observed $V_{ON}$ shifts in ISO-3 were approximately 10 times larger than those in ISO-10. Then, the trend of the shift in $V_{ON}$ observed could be proportional to the trap density at the semiconductor/insulator interface.

The activation energies of the charge carriers were investigated through temperature dependent electron transport to clarify the origin of the bias stress instability. Transfer curve measurements were carried out at different temperatures, ranging from 35 to 55 °C, with a $V_{DS}$ of 1 V. Figures 6(a) and (b) are the Arrhenius plots of $I_D$ vs $1/k_B T$ of ISO-3 and ISO-10 under various $V_{GS}$. Thermal activation of $I_D$ was clearly observed in the subthreshold region. Negative shifts of 5.6 and 2.0 V in $V_{ON}$ were observed for ISO-3 and ISO-10, respectively. The smaller shift in $V_{ON}$ suggests that higher $SiO_2$ incorporation suppressed the charge carrier excitation under thermal stresses. The $I_D$ obeyed the Arrhenius relationship as follows:

$$I_D = I_{D0} \exp\left(-\frac{E_a}{k_B T}\right), \qquad (5)$$

where $I_{D0}$ is the prefactor and $E_a$ is the activation energy for charge carriers. The $I_{D0}$ were almost constant irrespective of $V_{GS}$ above the threshold regions.[99] In general, $I_{D0}$ and $E_a$ are correlated in disordered systems. The relationship is given by the Meyer-Neldel (MN) rule as follows:[100]

$$I_{D0} = I_{D00} \exp\left(\frac{E_a}{E_{MN}}\right), \qquad (6)$$

where $E_{MN}$ is the MN energy. $I_{D0}$ and $E_a$ can be extracted from the Arrhenius plots. $E_a$ as a function of $V_{GS}$ drastically increased in the subthreshold region. $E_a$ corresponds to the difference between the Fermi level $E_F$ and the conduction band edge $E_C$ at absolute zero. $E_F$ can be tuned in TFTs by changing $V_{GS}$. The ISO-10 showed a larger $E_F$ shift than ISO-3 by sweeping small $V_{GS}$. Because the dependency of the $E_F$ shift reflects the density of states (DOS), when the $E_F$ reaches the band tail state near $E_C$, the shift is suppressed due to the exponential increase in DOS at the band tail.[101, 102] The large shift of $E_a$ in ISO-10 by sweeping small $V_G$ is evidence of a small DOS in the band tail. The $E_{MN}$ values estimated from the inverse of the slopes at high $E_a$ (subthreshold region of $V_{GS}$) were 30.0 and 29.2 meV for ISO-3 and ISO-10, respectively. These MN energies are quite comparable to those observed in IGZO[101, 103] and $a$-Si:H.[104] This universal $E_{MN}$ indicates that the electron transport at the conduction band tail could be expressed by the multiple trapping model.[105] Although the physical meaning of $E_{MN}$ is still a topic of discussion,[106] Mao $et\ al.$ proposed the following equation to explain this phenomenological model:

$$E_{MN} = \frac{E_a}{T_e/T_l - 1},  \qquad (7)$$

where $T_e$ and $T_l$ are the electron temperature and lattice temperature (device temperature), respectively.[107] Generally, under thermal equilibrium conditions where $T_e \approx T_l$, $E_{MN}$ and $E_a$ are strongly correlated for room temperature measurements. Comparing the ISO-3 with the ISO-10, smaller $E_a$ was observed for the ISO-10 in the regime above the threshold.[19] This result is consistent with the difference between the $E_{MN}$ of ISO-3 and ISO-10. At above threshold voltages, the linear relationship between $I_{D0}$ and $E_a$ disappeared. When $E_a$ was very small in ISO-10, the slope of $I_{D0}/E_a$ became negative, and $E_{MN}$ was also negative. Negative $E_{MN}$ were reported at low $E_a$ in IGZO TFTs and heavily doped microcrystalline-Si, where $E_F$ goes deeply into the band tail state.[101, 103, 104] This phenomenon is the so-called inverse MN rule. The inverse MN rule can only be seen in a system whose contact resistance is sufficiently low.[108] The inverse MN rule observed here also indicates a reasonably low ohmic contact.

To promote a deep understanding of the stabilities in ISO-3 and ISO-10, the DOS were extracted from capacitance-voltage measurements using a precision LCR meter (Agilent E4980A). AC voltages of 0.5 V and of 20 Hz were superimposed on DC gate voltages ($V_{GS}$). A schematic of the capacitance measurements is shown in the inset of Fig. 7, where $C_{i(D+S)}$ is the insulator capacitance between the back-gate and the drain to source electrodes. The amount of induced charge ($Q$) by $V_{GS}$ is given by the following:

$$Q = C(V_{GS} - V_{FB} - \phi_S), \qquad (8)$$

where $C \sim 120$ pF is the geometrical capacitance of the present TFTs, $V_{FB}$ is the flat band voltage, and $\phi_S$ is the electric potential of the ISO/SiO$_2$ surface. Eq. (8) can be transformed to the following:

$$\phi_S = \int_{V_{FB}}^{V_{GS}} \left(1 - \frac{C_{i(D+S)}}{C}\right) dV_{GS}. \qquad (9)$$

Energy band bending in semiconducting films becomes larger as the thickness of a film increases. ISO films 10 nm in thickness, in which the energy band bending was considered to be less than 50 meV,[109] gave sufficiently small gradients of electron concentrations. Then, it is assumed that the charge density ($\rho$) in the thin films were uniform. Using this assumption, $\rho$ as a function of $\phi_S$ was calculated by Eq. (8), and the DOS [$D(E)$] was extracted by the following:

$$D(E) = \left|\frac{d\rho}{d\phi_S}\right|_{\phi_S = E}, \qquad (10)$$

where the energy $E$ is defined by $E - E_C = -E_a$. The $D(E)$ calculated for ISO-3 and ISO-10 are shown in Fig. 7. The smaller $D(E)$ in ISO-10 beneath the $E_C$ indicates that the trap density in the film is smaller than that in ISO-3. The smaller DOS in ISO-10 at the deep energy level also supports the reduction in off-state current with increasing $SiO_2$ content [Fig. 4(d)].

## 5. Desorption of excess oxygen in vacuum storage conditions

### 5.1 TFT properties in atmosphere and vacuum

Although the ISO-10 TFTs demonstrate stable operation, the TFT mobility slightly deteriorates (7.7 cm$^2$/Vs). To maintain both the high mobility of the ISO-3 TFTs (18.4 cm$^2$/Vs) and electrical stability of the ISO-10, it is necessary to clarify the electronic conduction instability. Here, we consider the excess oxygen generated during sputtering deposition of an amorphous oxide semiconductor film. The excess oxygen is an important part of the composition in an $InO_x$-based film because excess oxygen can easily be incorporated into a film or easily diffuse out of the film, and the resulting TFT behaviors are drastically different.

We observed the effect of storage environment on the electrical stability to compare the excess oxygen contents in the ISO-3 and ISO-10 TFT. To adjust the $V_{ON}$ to be 0 V, the ISO-3 and ISO-10 films were deposited at high $P_{O2}$ [$O_2$/(Ar + $O_2$) = 50 %] and at low $P_{O2}$ [$O_2$/(Ar + $O_2$) = 8.3 %], respectively. The $V_{ON}$ is a key factor for TFT operation at low voltages. Due to the $P_{O2}$ optimization, the films with different $O_2$ incorporation are compared because a vast amount of oxygen molecules could be incorporated in the film under a high $O_2$ concentration.[51, 110, 111] Figure 8 shows a flow of the fabrication of the TFT and *I-V* characterization for the excess oxygen characterization. The TFT fabrication process was the same as that of the TFTs in the previous section. The TFTs fabricated were sequentially measured in different three states: immediately after fabrication (as-fabricated), after storage in a vacuum desiccator (~10 Pa) for 3 months after the first measurement (stored in vacuum), and after exposure to ambient air for 2 weeks after the second measurement (exposure to air).

Figure 9 shows the transfer characteristics of the ISO-3 TFTs. For the as-fabricated TFTs, transfer characteristics at different channel lengths ($L = 50−350$ μm in steps of 50 μm) are presented in Fig. 9(a). The maximum $I_D$ increased with decreasing $L$, while the $V_{ON}$ was almost constant. After being stored in a vacuum desiccator, however, the $V_{ON}$ significantly shifted to more negative voltages in the shorter channel TFTs. Even when $V_{GS} = −100$ V was applied, $I_D$ was not fully turned off [Fig. 9(b)]. After being exposed to air for two weeks, the $V_{ON}$ in each $L$ tended to return to zero [Fig. 9(c)]. Figures 9(d)-(f) are the transfer characteristics of ISO-10 TFTs with the same measurement conditions as those of the ISO-3 TFTs. In contrast to the large shift of $V_{ON}$ in the ISO-3 TFTs, no $V_{ON}$ shift was observed in the ISO-10 TFTs for any of the storage environments. For the TFTs stored in vacuum, the $I$-$V$ characteristics for the ISO10 films were significantly different than those for the ISO3 films. To observe the possible changes in the TFT properties after vacuum storage, we extracted the contact resistance ($R_c$) and channel resistivity ($r_{ch}$) of the TFTs using the transfer line method (TLM) as shown in Fig. 10. Note that the $V_{GS}$ is modified by the threshold voltage ($V_{th}$) in the ISO-3 TFTs because $V_{th}$ for each $L$ is different [Fig. 9(b)]. The width-normalized total resistance ($R_{total}W$) as a function of $L$ is expressed by the following:[56, 112]

$$R_{total}W = r_{ch}LW + R_cW, \qquad (11)$$

where $W$ is the channel width, which was fixed at 1000 μm. $R_c$ considers the contact resistance at the source and drain electrodes ($R_c = R_{source} + R_{drain}$). The $R_c$ extracted, which is determined by the vertical axis of the intersection points for each line, increased for the ISO-3 TFTs that were exposed to air. The increase in $R_c$ corresponds to an increase in the potential barrier height at the interface between the oxide semiconductors and contact electrodes due to the adsorbed oxygen molecules.[113-115] Since the absorbed $O_2$ causes charge transfer from the ISO film, barriers are generated due to surface band bending.[49] In contrast, as shown in Figs. 9(d)-(f), the $R_c$ of the ISO-10 TFTs are almost constant for each state. In particular, an $R_c$ higher than that of ISO-10 was observed in the ISO-3 TFTs that were after exposed to air. The high $R_C$ observed confirmed that the contact resistance was dominated not only by the interfacial oxidation status[56, 116] but also by the density of states of the thin film channel, as shown in Fig. 7.[117, 118]

## 5.2 Change in the carrier transport mechanism

To investigate the intrinsic film conductivity, we discuss the intrinsic mobility ($\mu_i$) using the following basic equation:[57]

$$\mu_i = \frac{1}{C_i W} \frac{\partial \sigma_{ch}}{\partial V_{GS}}, \qquad (12)$$

where $C_i$ is the gate insulator capacitance per unit area (13.8 nF/cm$^2$) and $\sigma_{ch}$ is the channel conductivity ($\sigma_{ch} = 1/r_{ch}$). Figure 11 shows the reciprocal $r_{ch}$ as a function of $V_{GS} - V_{th}$ for the ISO-3 TFTs, and the reciprocal $r_{ch}$ as a function of $V_{GS}$ for the ISO-10 TFTs in each state. The $\mu_i$ obtained for the ISO-3 were 18.2, 21.9, and 18.5 cm$^2$/Vs for the as-fabricated TFTs, TFTs stored in vacuum, and TFTs exposed to air, respectively [Fig. 11(a)], whereas a constant $\mu_i$ of ~7.0 cm$^2$/Vs was obtained for ISO-10 in all cases [Fig. 11(b)].[119] These results indicate that the TFTs become environmentally stable when using ISO-10 films sputtered with low $P_{O2}$. Compared with the ISO-3 TFTs, the reliability of the ISO-10 TFTs was significantly improved, although the $\mu_i$ was reduced.

To obtain more detail regarding the conduction change in the ISO-3 TFTs after vacuum storage, the Fermi level energy ($E_F$) was extracted by the following:[120]

$$E_F - E_m = \frac{k_B T}{q} \ln\left[\frac{n}{N_c} + \frac{1}{2^{3/2}}\left(\frac{n}{N_c}\right)^2\right], \qquad (13)$$

where $E_m$ is the mobility edge and $N_c$ is the effective density of states, which is estimated using an electron effective mass of ~0.35 $m_e$ ($m_e$ is the electron rest mass) for amorphous InO$_x$-based semiconductors.[3] The carrier density ($n$) as a function of $V_{GS}$ [$n(V_{GS})$] was extracted using $\mu_i$ based on the following standard scheme:[121]

$$n(V_{GS}) = \frac{\sigma_{ch}}{q\mu_i} = \frac{1}{q\mu_i}\frac{\partial I_D(V_{GS})}{\partial V_{DS}}\frac{L}{Wt}, \qquad (14)$$

where $t$ is the channel thickness. As shown in Fig. 12(a), $E_F$ resides within the localized tail states for the as-fabricated ISO-3 TFTs. InO$_x$ thin films generally have an exponential DOS at the conduction band edge (band tail states).[102] The Fermi level shifts upwards under application of a positive $V_{GS}$. The DOS is exponentially increased and is filled by electrons, where the charge carrier transport is described by hopping mechanisms, i.e., so-called trap-limited conduction. On the other hand, for ISO-3 TFTs stored in vacuum, $E_F$ is between $E_m$ and $\phi_0$, where the $\phi_0$ is the average height of the potential barriers. The average of the barrier heights is within the range of 40−120 meV.[122] The discussion above potentially suggests that the carrier transport is strongly dominated by percolation conduction.[123, 124]

Excess oxygen could form deep subgap states below conduction band edges[109] that act as electron traps[110] because of the charge transfer from the conduction paths of the In ions.[7] Then, the excess oxygen could be removed by low-temperature annealing in a reducing atmosphere.[125] An X-ray absorption near-edge structure (performed at the BL27SU beamline of the SPring-8 synchrotron radiation facility) spectrum demonstrated that the coordination number of the Si atoms in the films was four under a vacuum of $10^2$ Pa.[87] Since it was also confirmed that Si-dopants tightly hold oxygen atoms around them, the excess oxygen is nonbonded (or weakly bonded) to the Si atoms in the film. Thus, we can infer that the molecules desorbed under vacuum could be excess oxygen, as illustrated in the inset of Fig. 12(a). The oxygen removal from the film causes an increase in carrier density and leads to the shift of $E_F$ above $E_m$. Then, the electron transport changes from trap-limited to percolation conduction. In contrast, adsorbing oxygen from the ambient atmosphere reduces carrier concentration due to the charge transfer.[71, 114, 126-128] The effect of oxygen adsorption is also reflected in the shift in $E_F$ from above to below $E_m$, as shown in Fig. 12(b), and then, the $I$-$V$ characteristics of the vacuum storage TFTs are recovered. This situation is dissimilar from the dense packing of polycrystalline ISO, which hinders the reabsorption of weakly bonded oxygen in the film.[87] Furthermore, in the percolation model, carrier transport strongly depends on the channel dimensions ($L$, $W$ and $t$). The $V_{th}$ is also affected by these dimensions.[129, 130] For longer channel TFTs, a higher $V_{GS}$ is required to achieve a certain electrical conductivity compared with that in shorter channels. This phenomenon is consistent with the $I$-$V$ characteristics observed for the TFT stored in vacuum [Fig. 9(b)].

## 6. Bilayer channel TFT

Although $V_O$ is necessary for electrical conduction in an $InO_x$-based semiconductor, the density of $V_O$ can be easily changed by adsorption/desorption of oxygen and/or hydrogen, which drastically degrades TFT properties. Furthermore, hydrogen contamination in oxide films may also cause instability in TFT operation.[7, 48, 126, 127, 131] To avoid these gas sensitivities, we propose a homogeneous bilayer oxide semiconductor channel, where the top layer can prevent gas diffusion in and out of the film. The concept of the bilayer channel is schematically illustrated in Fig. 13(a). Bottom-gate top-contact TFTs were fabricated on a heavily doped p-type Si substrate with 250 nm of thermally grown $SiO_2$. The bilayer ISO films were sequentially deposited using DC magnetron sputtering without exposure to atmosphere. The sputtering targets for the ISO contained 3 and 20.6 wt. % $SiO_2$ and are denoted ISO-3 and ISO-20, respectively. The active channel was 5-nm-thick stacked ISO-3 and ISO-20. Because the Ti electrodes were directly evaporated on the bilayer, the Ti electrodes were located on the ISO-20 insulator without forming a hole [Fig. 13(b)]. The fabricated TFTs were finally annealed at 250 °C for 30 min in air. After annealing the stacked TFTs, we confirmed that the bilayer channel (ISO-3/20) was amorphous by using cross-section TEM [Fig. 13(c)] and selected area diffraction, which indicates a series of concentric diffuse rings [Fig. 13(d)]. Using X-ray diffraction (BL04B2 beamline at the SPring-8 synchrotron radiation facility) and X-ray absorption fine structure measurements (BL01B1 beamline at the SPring-8 synchrotron radiation facility), we confirmed that all of the films prepared in the same manner remained amorphous.

The bilayer TFT with a channel length of 350 μm and channel width of 1000 μm showed no pronounced hysteresis in its transfer [Fig. 14(a)] and output [Fig. 14(b)] characteristics. Though the contact was formed on the insulating ISO-20 layer, no current crowding effect was observed in the I–V characteristics in the low-bias regime, indicating the formation of ohmic contact. A cross-sectional TEM image of the Ti contact reveals that the Ti at the ISO-20 interface is clearly oxidized.[116, 132] This result implies that the Ti contact takes oxygen molecules from the ISO-20 film, forming $V_O$ in the ISO-20. Through this process, the thin ISO-20 becomes conductive underneath the Ti contact, even though an insulating top layer was used. Carrier injection without any holes is fascinating due to process simplification, and the bilayer structure is unlike a TFT with an etched stop layer. The excess oxygen in the ISO-20 causes oxidation of the Ti interface because of the low Gibbs free energy of Ti (–848 kJ/mol at 250 °C).[133] The results are consistent from the viewpoint of oxygen bindings.

The TFT parameters of $V_{ON}$, $\mu_{FE}$, $I_{ON}/I_{OFF}$, and $ss$ obtained are –0.4 V, 19.6 cm$^2$/Vs, 1.0 × 10$^8$, and 0.1 V/dec, respectively. We emphasize that a TFT was successfully achieved with both high mobility (> 10 cm$^2$/Vs) and normally off operation even though the metallic contacts were formed on the insulating ISO-20 layer covering the semiconducting ISO-3. The single-layer ISO-3 TFT shows a large negative shift in $V_{ON}$, and as a result, an unintentionally high carrier density is produced. Thus, the normally off characteristics indicate the effective suppression of oxygen desorption from the bottom layer of the ISO-3/20 channel. Similar effects have been reported for other insulator oxide films,[134, 135] in which the insulating film covering the oxide semiconducting channel suppresses the $V_{ON}$ shift. In the bilayer TFT structure, the insulative ISO-20 can act as a protection layer for the adsorption/desorption of oxygen atoms to/from the semiconducting bottom layer without intentionally forming contact holes.

Finally, we mention the bias stress instability of the ISO-3/20 TFT. Figure 15(a) and (b) show the negative ($V_{GS} = $ –20 V, 5000 s) and positive ($V_{GS} = $ +20 V, 5000 s) bias stress instability measurements, respectively. Although the $V_{GS}$ applied were different from those of the single-layer ISO-10 TFTs (Fig. 5), the $V_{ON}$ shifts of 0.2 and 0.8 V for NBS and PBS, respectively, are very small, similar to crystalline IGZO TFT.[136] Significant improvement of the stability while maintaining high mobility and normally off operation can be achieved in the homogeneous ISO stacked channel. This result implies that $V_o$ formation, which the primary origin of the bias stress instability,[137] is suppressed because the top layer prevents adsorption/desorption of $O_2$ molecules on/from the bottom layer. In addition, the thin insulating ISO-20 does not behave as a barrier for ohmic contact at the Ti electrode interface. Here, we emphasize that the unique and reliable bilayer ISO TFTs will allow these structures to be used in practical applications.

# 7. Conclusions

We  have discussed the bond-dissociation energies of dopants in $InO_x$-based semiconductors for controlling $V_O$ formation. We compared three dopants (Ti, W, and Si) in $InO_x$ films. These species have different BDE; thus, the choice of dopant modulated the electrical properties of the TFTs. Si dopant in $InO_x$ can realize higher carrier density and field effect mobility than W and Ti dopants. Furthermore, the electrical properties could then stabilize when Si concentration increases. In particular, ISO-10 showed stable TFT operation even in the single-layer TFT structures due to the $V_O$ suppression and reduction in the DOS beneath the conduction band. We clarified that the instability of the ISO-3 is due to the desorption of excess oxygen in the film; however, the TFT with low Si concentration exhibited high mobility. We demonstrated bilayer ISO TFTs that realize high mobility (19.6 cm$^2$/Vs), high stability under bias stress instability measurements, and normally off operation. ISO TFTs are expected to lead to next-generation applications as representatives of the postamorphous

Si industries.

## Acknowledgments

The authors would like to thank Dr. N. Mitoma (Nagoya University) and Dr. T. Kizu (K.K. Air Liquide Laboratories) for collaborative fabrications and measurements with fruitful discussions for understanding ISO properties. The authors also acknowledge Dr. X Gao (Soochow University, China), Dr. M. Shimizu (Tokyo University of Science), Dr. M.-F. Lin (Cranfield University, UK), Dr. W. Ou-Yang (East China Normal University, China), Prof. A. Fujiwara (Kwansei Gakuin University), Dr. Ina (Japan Synchrotron Radiation Research Institute: JASRI), Dr. Y. Tamenori (JASRI), Dr. T. Uruga (JASRI), and K. Yanagisawa (RIKEN) for valuable experiments and important suggestions regarding the oxide films and K. Ohno and K. Tamura (NIMS) for their kind help regarding film formations. The authors thank Sumitomo Metal Mining Co., Ltd. for supplying the ITiO and IWO sputtering targets. This work was partially supported by Grants-in-Aid for Scientific Research (Grant Nos. 15H03568, 26790051, and 18K18868) and the Open Partnership Joint Project of JSPS-NSFC Bilateral Joint Research (Grant No. 61511140098). The synchrotron radiation experiments were performed with the approval of the Japan Synchrotron Radiation Research Institute (Proposal Nos. 2015A1884 and 2015A1885). Part of this work was performed under the Cooperative Research Program of the Institute for Joining and Welding Research at Osaka University.

## Figure Captions

**Fig. 1.** (a) Schematic diagram of the device structure. The p$^{++}$-Si substrate acts as a back-gate electrode. The inset shows an AFM image of the postannealed IWO channel. The estimated root-mean-square roughness is 0.27 nm. Typical *I-V* characteristics of postdeposition annealed IWO TFT: (b) output and (c) transfer characteristics. The channel length and width of the device were 350 and 1000 μm, respectively. TFT characterization was carried out at room temperature in the dark under vacuum conditions (~10$^{-4}$ Pa). Adapted with permission from Ref. 22. Copyright 2013. AIP Publishing LLC.

**Fig. 2.** Electrical conductivity of as-deposited films as a function of the oxygen partial pressure used during sputtering. The film conductivity was extracted from the linear region of the output characteristics at $V_{GS}$ = 30 V. The inset shows schematic images of the films doped with high and low BDE, which is related to the change in the conductivity. Adapted with permission from Ref. 18. Copyright 2013. AIP Publishing LLC.

**Fig. 3.** Typical output and transfer characteristics of fabricated TFTs: (a, b) ITiO, (c, d) IWO, and (e, f) ISO. The devices were annealed at 150 °C in air. The TFTs were measured at room temperature in the dark under ambient atmosphere. The TFT dimensions include a channel length *L* of 350 μm, a channel width *W* of 1000 μm, and a gate capacitance per unit area $C_i$ of 1.21×10$^{-8}$ F/cm$^2$ (based on a dielectric constant of 3.9 for SiO$_2$). Adapted with permission from Ref. 18. Copyright 2013. AIP Publishing LLC.

**Fig. 4.** Typical transfer characteristics of the ISO-3 (a) and ISO-10 (b) TFTs. Electrical properties of ISO TFTs over various SiO$_2$ contents and oxygen partial pressures during sputtering, showing changes in (c) $\mu_{FE}$, (d) $I_{OFF}$ and $I_{ON}$ (blue) (e) $ss$, and (f) $V_{ON}$. The dotted lines are guides for the eyes. The electrical measurements were performed at room temperature in the dark under high vacuum (~$10^{-4}$ Pa). Adapted with permission from Ref. 19. Copyright 2014. AIP Publishing LLC.

**Fig. 5.** Comparison of $V_{ON}$ shift when applying negative (a) and positive (b) gate biases for ISO-3 and ISO-10 TFTs. The dotted lines are guides for the eyes. Adapted with permission from Ref. 19. Copyright 2014. AIP Publishing LLC.

**Fig. 6.** Temperature dependence of $I_D$ at various $V_{GS}$ for (a) ISO-3 and (b) ISO-10 at $V_{DS} = 1$ V. The closed circles denote measurements from the subthreshold region. The solid lines are least-squares fits to the experimental data using an Arrhenius relation. Adapted with permission from Ref. 19. Copyright 2014. AIP Publishing LLC.

**Fig. 7.** Calculated DOS beneath the conduction band edge. The red and black circles are data extracted from ISO-3 and ISO-10, respectively.

**Fig. 8.** Fabrication and measurement flow of ISO TFTs with low (ISO-3) and high (ISO-10) $SiO_2$ concentration channels. The oxygen concentrations are optimized during sputtering of ISO-3 and ISO-10 channels by adjusting the $V_{ON}$ to be approximately 0 V. The inset shows a schematic diagram of the device configuration.

**Fig. 9.** Comparison of transfer characteristics of TFTs using ISO-3 (a, b, c) and ISO-10 (d, e, f) conditions with different channel lengths (50 – 350 μm in steps of 50 μm) for the as-fabricated TFTs (a, d), TFTs stored in vacuum (10 Pa) for 3 months (b, e), and TFTs that were then exposed to air for 2 weeks (c, f). Adapted with permission from Ref. 119. Copyright 2015. AIP Publishing LLC.

**Fig. 10.** Width-normalized total resistance ($R_{total}W$) extracted at $V_{DS}$ of 1 V as a function of $L$. The $V_{GS}$ is modified by $V_{th}$ for the ISO-3 TFTs (a, b, c) since $V_{th}$ was drastically changed after storage in vacuum conditions. On the other hand, as for the ISO-10 TFTs (d, e, f), $V_{th}$ was independent of the storage conditions. The inset shows the magnification of the intersection point. Adapted with permission from Ref. 119. Copyright 2015. AIP Publishing LLC.

**Fig. 11.** (a) Reciprocal $r_{ch}$ as a function of $V_{GS} - V_{th}$ for the ISO-3 TFTs. (b) The reciprocal $r_{ch}$ as a function of $V_{GS}$ for the ISO-10 TFTs. Adapted with permission from Ref. 119. Copyright 2015. AIP Publishing LLC.

**Fig. 12.** Fermi energy as a function of $V_{GS}$ for the ISO-3 TFTs: (a) stored in vacuum and (b) exposed to air. The Fermi energy, plotted in black, was measured in the initial as-fabricated TFT. After storing the TFT in vacuum, the energy (plotted in red) was measured. Then, after exposure to air, the Fermi energy changed to the line plotted in blue. Above and below $E_m$ correspond to percolation and trap-limited conduction, respectively. The inset shows a schematic illustration of the oxygen desorption/adsorption model. Adapted with permission from Ref. 119. Copyright 2015. AIP Publishing LLC.

**Fig. 13.** (a) Concept of a bilayer channel. The top layer prevents desorption of excess oxygen

in the bottom film. (b) Schematic diagram of the device structure. (c) Cross-sectional TEM image of the bilayer films with vertically stacked ISO-20 and ISO-3 on a $SiO_2$/Si substrate. (d) Electron-beam selected area diffraction image. Adapted with permission from Ref. 132. Copyright 2016. AIP Publishing LLC.

**Fig. 14.** (a) Transfer characteristics and (b) output characteristics of the bilayer TFT using ISO-3/20 stacked channel. Adapted with permission from Ref. 132. Copyright 2016. AIP Publishing LLC.

**Fig. 15.** Negative (a) and positive (b) gate bias stress instabilities measured using bilayer TFT. The bias stress was applied at $V_{GS} = \pm 20$ V for 5000 s. The device was measured in air in the dark.

**Table I.** Comparison of typical TFT properties estimated from the transfer characteristics presented in Fig. 3. $I_{ON}/I_{OFF}$ is defined as the ratio of the maximum $I_D$ to the minimum $I_D$ in the graph ($-20$ V $\leq V_{GS} \leq 40$ V). $V_{ON}$ is defined as the $V_{GS}$ at which $I_D$ begins increasing. Adapted with permission from Ref. 18. Copyright 2013. AIP Publishing LLC.

| | $I_{ON}/I_{OFF}$ | $ss$ (V/dec) | $V_{ON}$ (V) | $\mu_{sat}$ (cm$^2$/Vs) |
|---|---|---|---|---|
| ITiO | $9.4 \times 10^9$ | 0.30 | $-4.0$ | 32 |
| IWO | $5.2 \times 10^9$ | 0.46 | $-6.3$ | 30 |
| ISO | $4.8 \times 10^9$ | 0.29 | 0.0 | 17 |

**Figures**

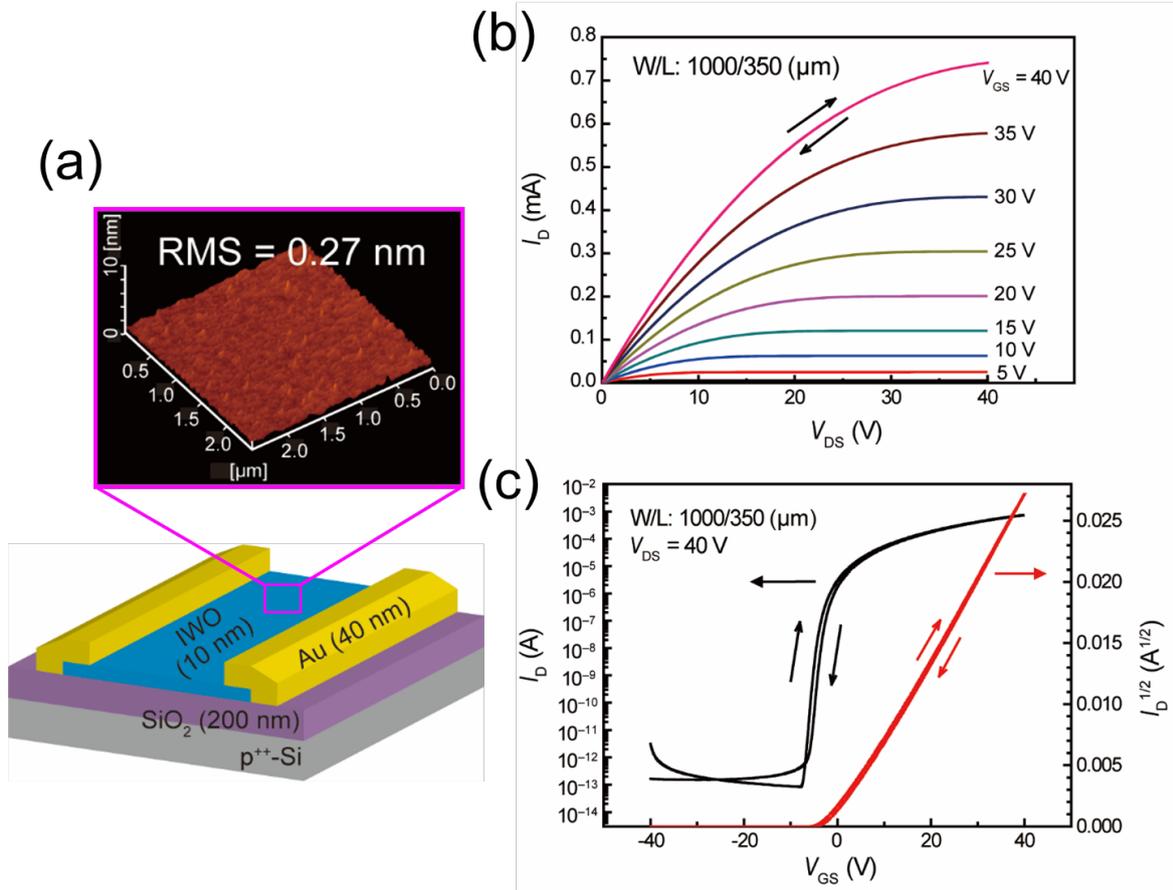

**Fig. 1.**

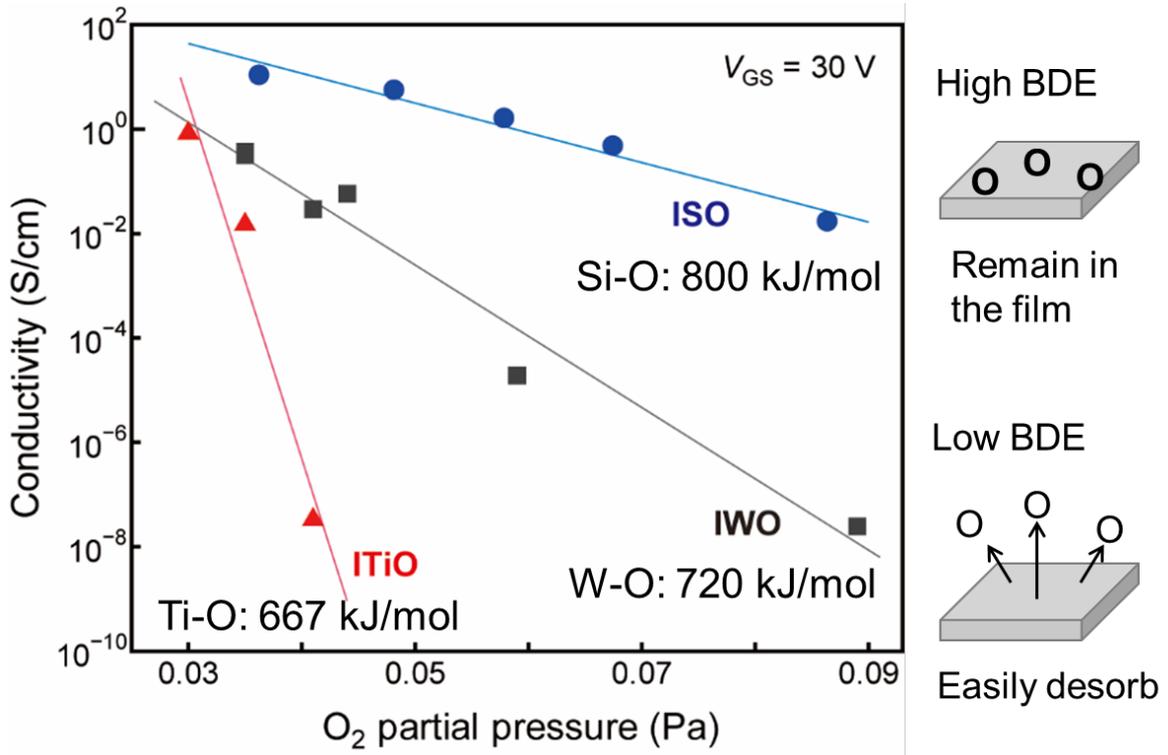

**Fig. 2.**

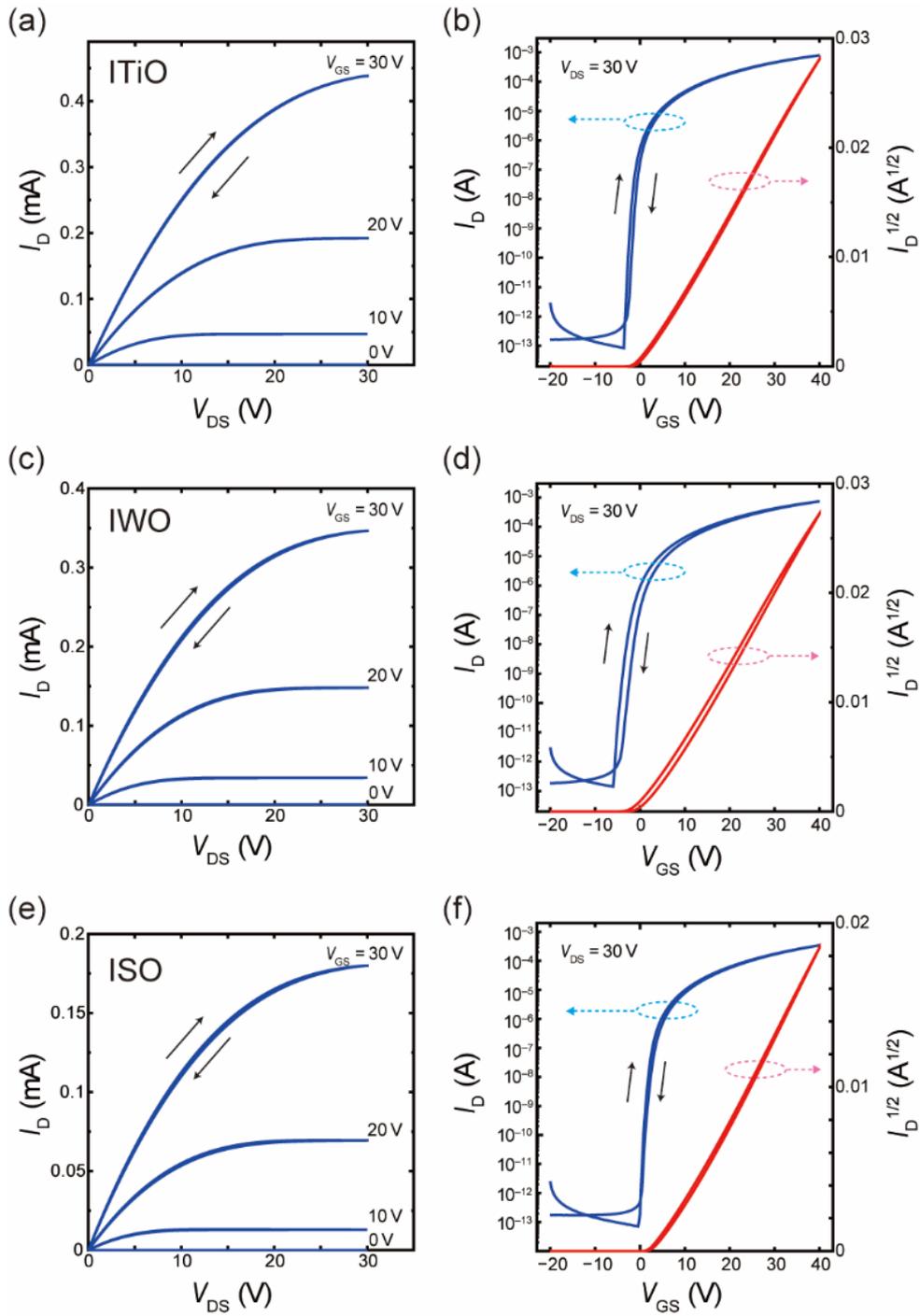

**Fig. 3.**

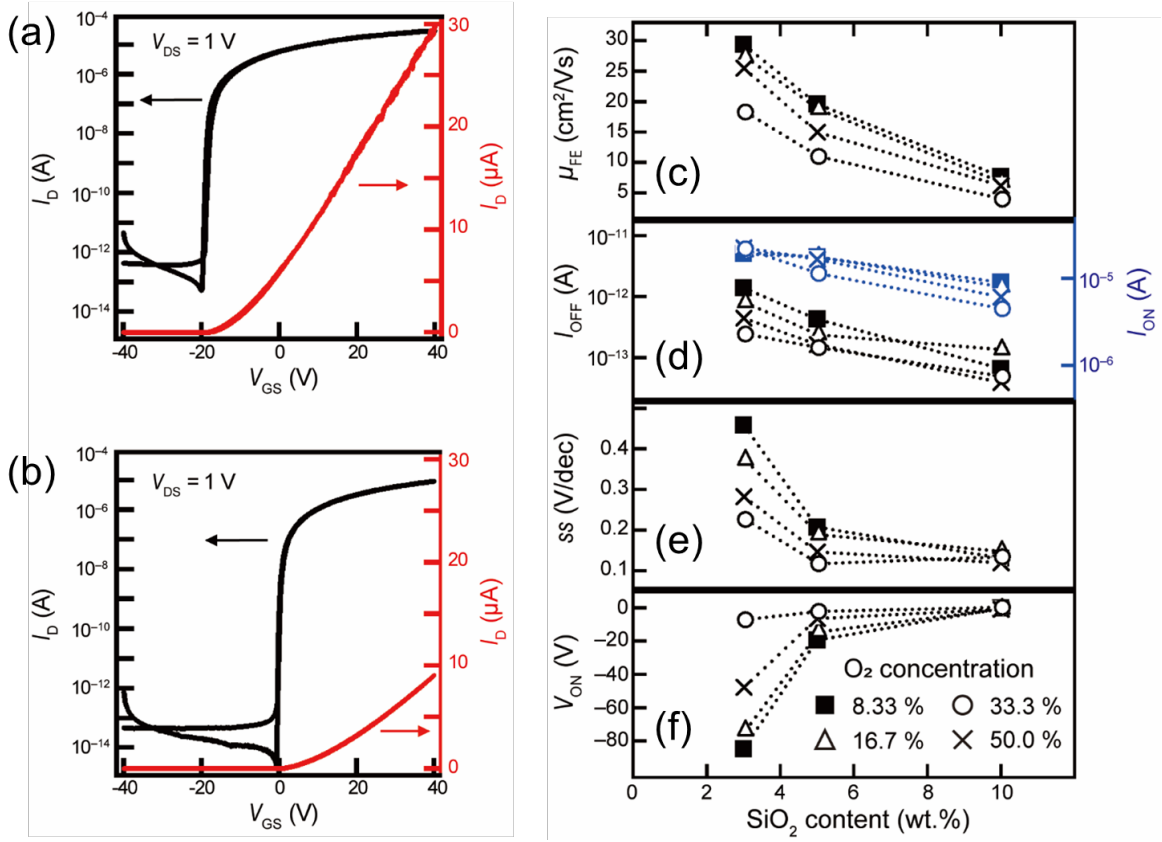

**Fig. 4.**

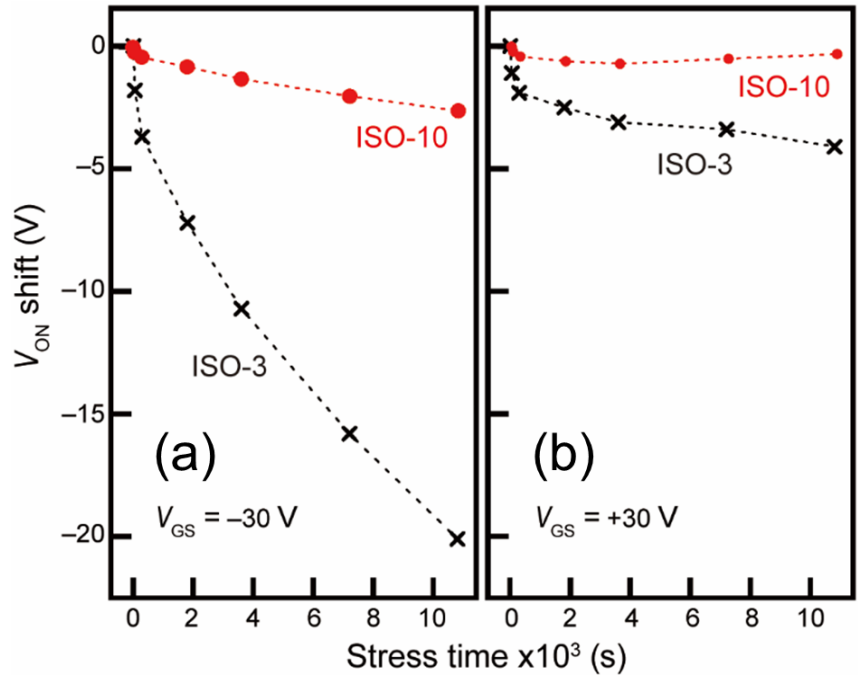

**Fig. 5.**

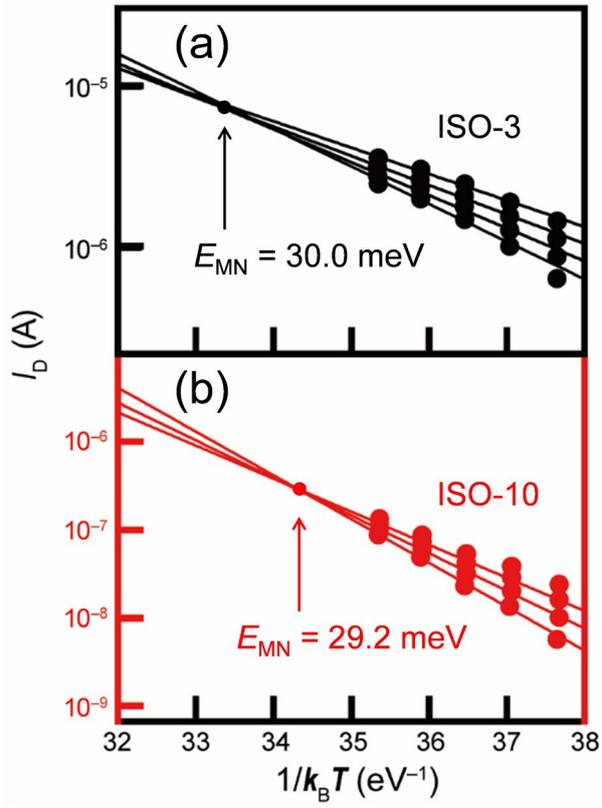

**Fig. 6.**

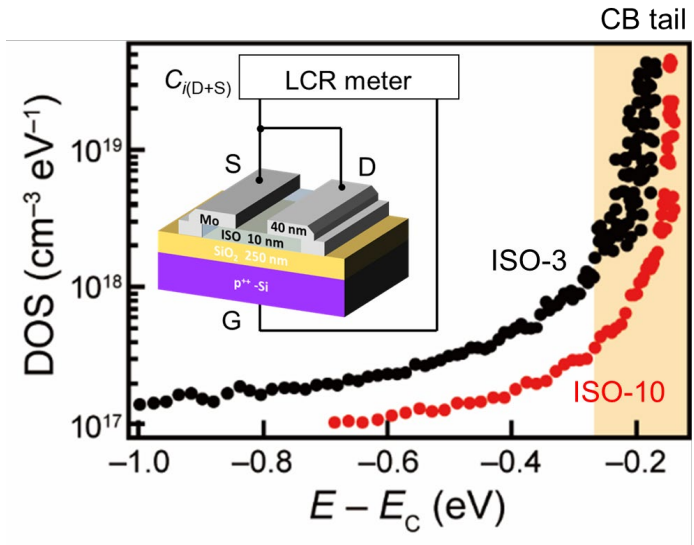

**Fig. 7.**

- ISO channel deposition
  DC power: 200 W, $P_{total}$: 0.25 Pa
  ISO-3: $O_2$ = 50%
  ISO-10: $O_2$ = 8.3%

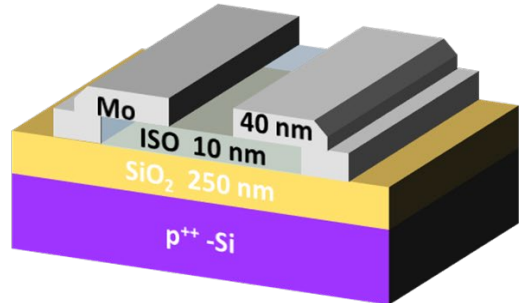

- Pre-annealing: 250 °C, 30 min in air

- Mo S/D electrode deposition

- Post-annealing: 150 °C, 30 min in $O_3$

- *I-V* characterization
  1st meas.: immediately after fabrication
  2nd meas.: after storage in a vacuum desiccator for 3 months
  3rd meas.: after exposure to ambient air for 2 weeks

**Fig. 8.**

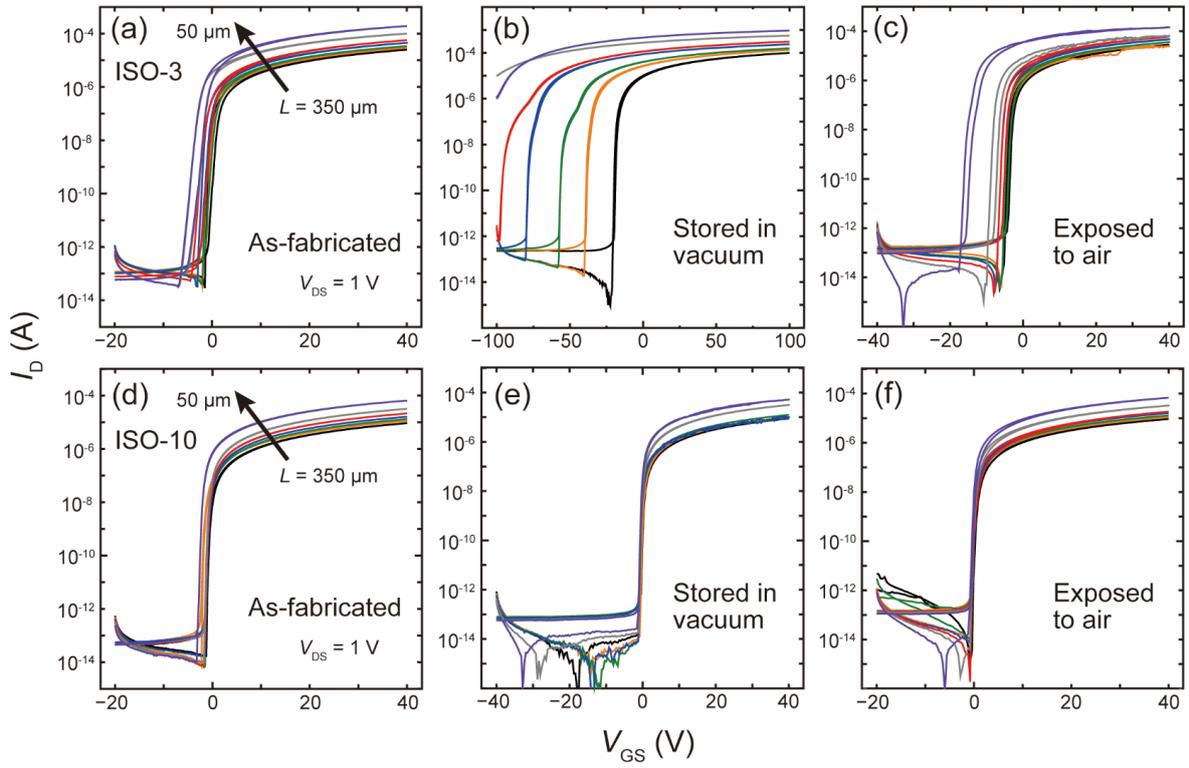

**Fig. 9.**

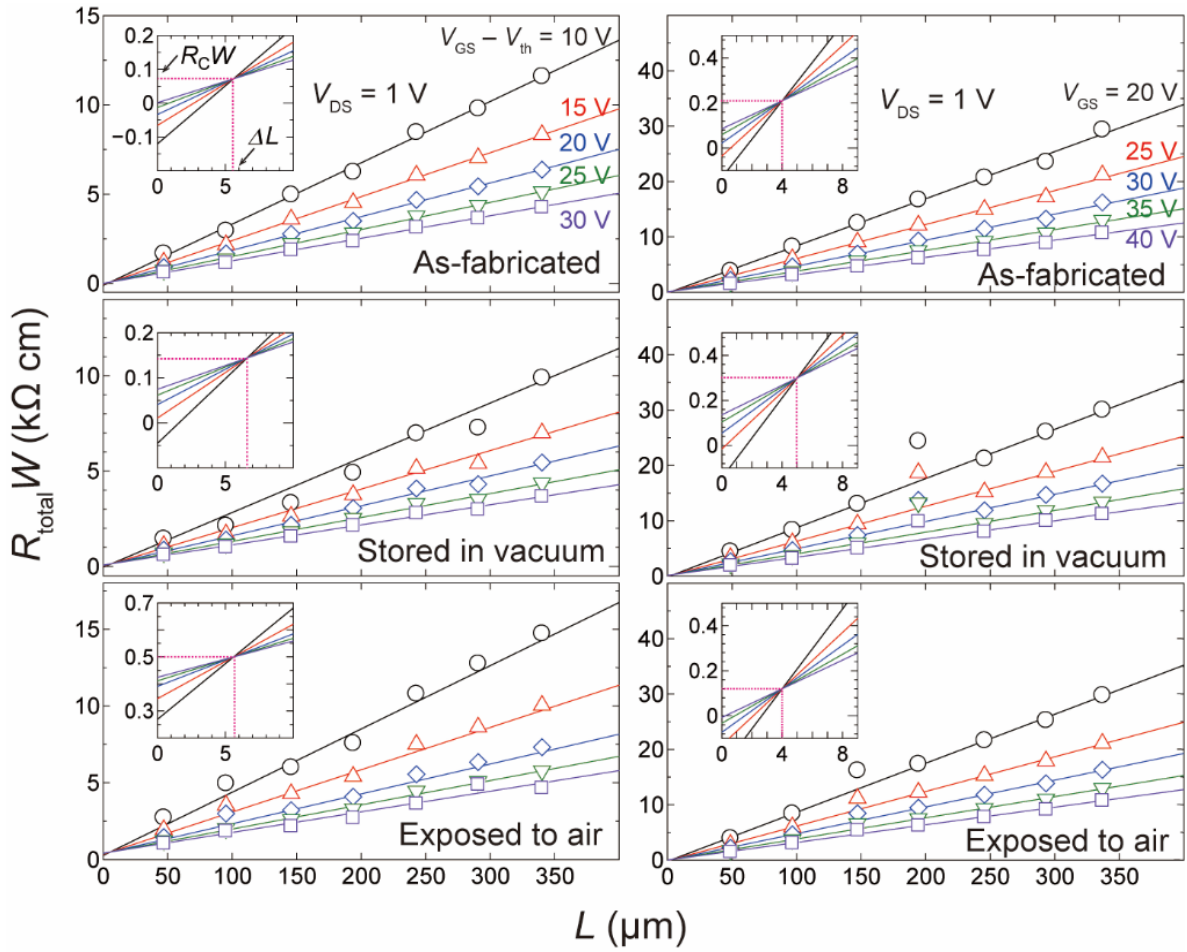

**Fig. 10.**

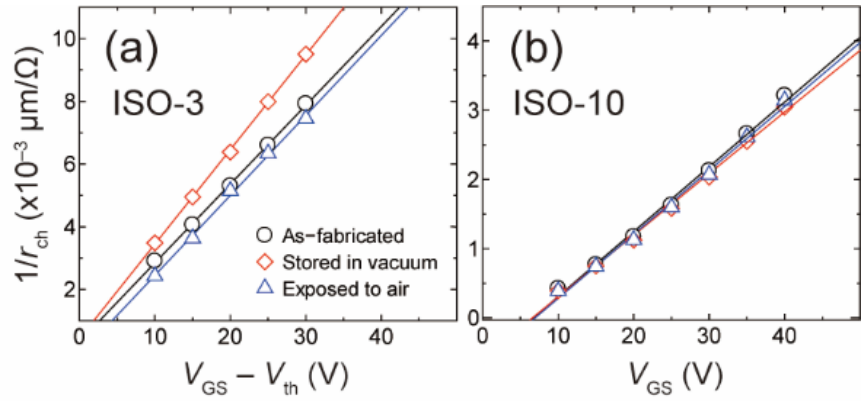

**Fig. 11.**

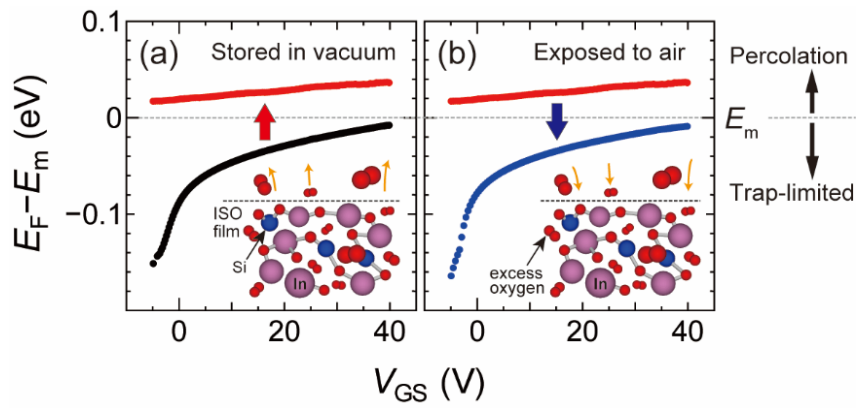

**Fig. 12.**

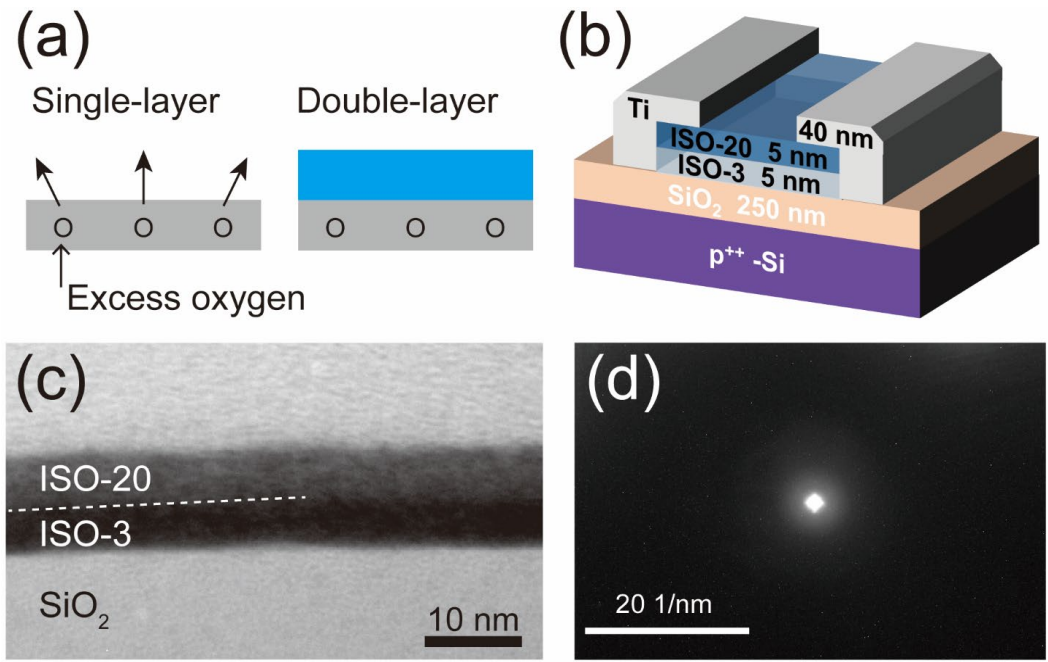

**Fig. 13.**

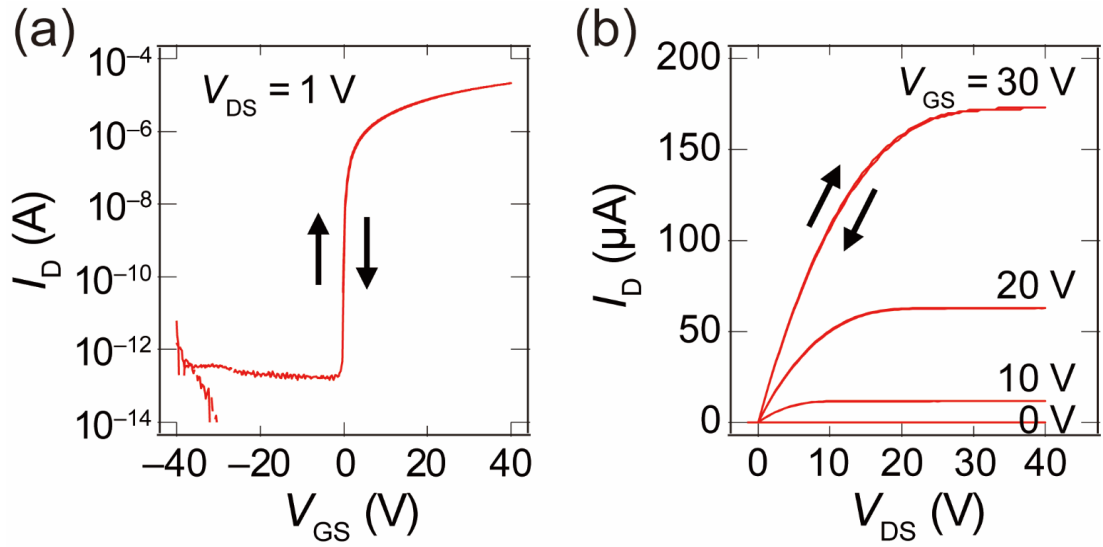

**Fig. 14.**

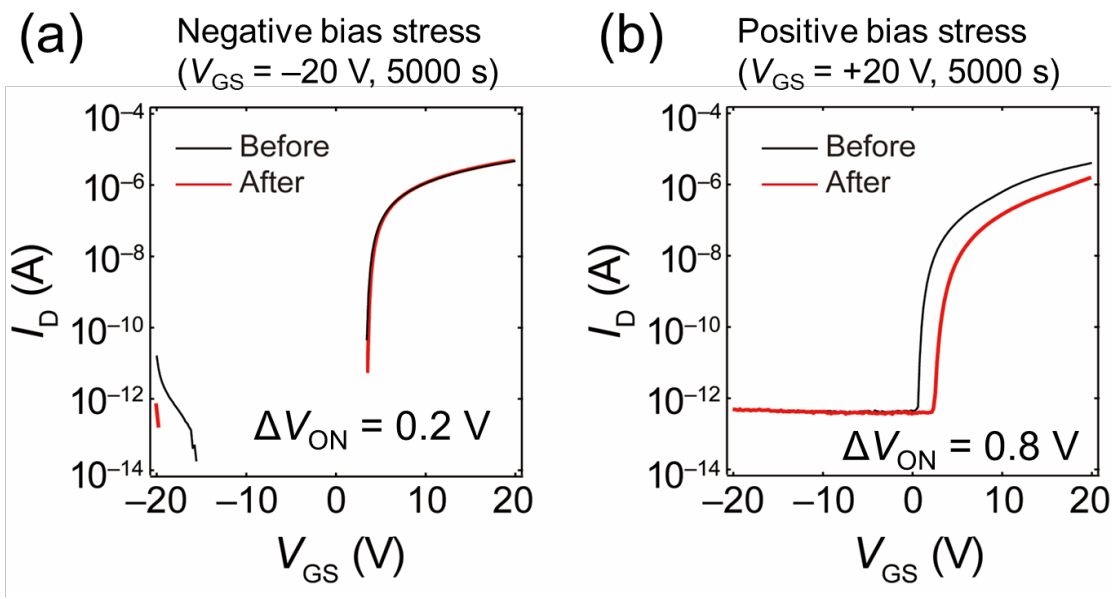

**Fig. 15.**

# References

1) H. Fritzsche and K. J. Chen, Phys. Rev. B **28**, 4900 (1983).
2) K. Nomura, H. Ohta, A. Takagi, T. Kamiya, M. Hirano, and H. Hosono, Nature **432**, 488 (2004).
3) T. Kamiya and H. Hosono, NPG Asia Mater. **2**, 15 (2010).
4) T. Kamiya, K. Nomura, and H. Hosono, Sci Technol Adv Mater **11**, 044305 (2010).
5) H. Hosono, N. Kikuchi, N. Ueda, and H. Kawazoe, J. Non-Cryst. Solids **198-200**, 165 (1996).
6) K. Nomura, T. Kamiya, H. Ohta, T. Uruga, M. Hirano, and H. Hosono, Phys. Rev. B **75**, 035212 (2007).
7) T. Kamiya, K. Nomura, and H. Hosono, J. Disp. Technol. **5**, 273 (2009).
8) K. Nomura, A. Takagi, T. Kamiya, H. Ohta, M. Hirano, and H. Hosono, Jpn. J. Appl. Phys. **45**, 4303 (2006).
9) J. Liu, D. B. Buchholz, R. P. Chang, A. Facchetti, and T. J. Marks, Adv. Mater. **22**, 2333 (2010).
10) Dhananjay and C.-W. Chu, Appl. Phys. Lett. **91**, 132111 (2007).
11) N. Joo Hyon, R. Seung Yoon, J. Sung Jin, K. Chang Su, S. Sung-Woo, P. D. Rack, K. Dong-Joo, and B. Hong Koo, IEEE Electron Device Lett. **31**, 567 (2010).
12) H. K. Müller, Phys. Status Solidi B **27**, 723 (1968).
13) H. Kumomi, S. Yaginuma, H. Omura, A. Goyal, A. Sato, M. Watanabe, M. Shimada, N. Kaji, K. Takahashi, and M. Ofuji, J. Disp. Technol. **5**, 531 (2009).
14) K. Ebata, S. Tomai, Y. Tsuruma, T. Iitsuka, S. Matsuzaki, and K. Yano, Appl. Phys. Express **5**, 011102 (2012).
15) G. Gonçalves, P. Barquinha, L. Pereira, N. Franco, E. Alves, R. Martins, and E. Fortunato, Electrochem. Solid-State Lett. **13**, H20 (2010).
16) T. Maruyama and T. Tago, Appl. Phys. Lett. **64**, 1395 (1994).
17) M.-F. Lin, X. Gao, N. Mitoma, T. Kizu, W. Ou-Yang, S. Aikawa, T. Nabatame, and K. Tsukagoshi, AIP Adv. **5**, 017116 (2015).
18) S. Aikawa, T. Nabatame, and K. Tsukagoshi, Appl. Phys. Lett. **103**, 172105 (2013).
19) N. Mitoma, S. Aikawa, X. Gao, T. Kizu, M. Shimizu, M.-F. Lin, T. Nabatame, and K. Tsukagoshi, Appl. Phys. Lett. **104**, 102103 (2014).
20) T. Miyasako, M. Senoo, and E. Tokumitsu, Appl. Phys. Lett. **86**, 162902 (2005).
21) S. Y. Park, K. H. Ji, H. Y. Jung, J.-I. Kim, R. Choi, K. S. Son, M. K. Ryu, S. Lee, and J. K. Jeong, Appl. Phys. Lett. **100**, 162108 (2012).
22) S. Aikawa, P. Darmawan, K. Yanagisawa, T. Nabatame, Y. Abe, and K. Tsukagoshi, Appl. Phys. Lett. **102**, 102101 (2013).
23) T. Kizu, S. Aikawa, N. Mitoma, M. Shimizu, X. Gao, M.-F. Lin, T. Nabatame, and K. Tsukagoshi, Appl. Phys. Lett. **104**, 152103 (2014).
24) N. L. Dehuff, E. S. Kettenring, D. Hong, H. Q. Chiang, J. F. Wager, R. L. Hoffman, C. H. Park, and D. A. Keszler, J. Appl. Phys. **97**, 064505 (2005).
25) P. Barquinha, A. Pimentel, A. Marques, L. Pereira, R. Martins, and E. Fortunato, J. Non-Cryst. Solids **352**, 1749 (2006).
26) S. Parthiban and J.-Y. Kwon, J. Mater. Chem. C **3**, 1661 (2015).
27) K. Kurishima, T. Nabatame, N. Mitoma, T. Kizu, S. Aikawa, K. Tsukagoshi, A. Ohi, T. Chikyow, and A. Ogura, J. Vac. Sci. Technol., B **36**, 061206 (2018).
28) K. Kurishima, T. Nabatame, T. Kizu, N. Mitoma, K. Tsukagoshi, T. Sawada, A. Ohi, I. Yamamoto, T. Ohishi, T. Chikyow, and A. Ogura, ECS Trans. **75**, 149 (2016).
29) J.-Y. Kwon and S. Parthiban, RSC Adv. **4**, 21958 (2014).
30) K. Ghaffarzadeh, A. Nathan, J. Robertson, S. Kim, S. Jeon, C. Kim, U. I. Chung, and J.-H. Lee, Appl. Phys. Lett. **97**, 143510 (2010).
31) J. W. Hennek, J. Smith, A. Yan, M. G. Kim, W. Zhao, V. P. Dravid, A. Facchetti, and T. J.

Marks, J Am Chem Soc **135**, 10729 (2013).

32) W. Ou-Yang, N. Mitoma, T. Kizu, X. Gao, M.-F. Lin, T. Nabatame, and K. Tsukagoshi, Appl. Phys. Lett. **105**, 163503 (2014).

33) E. Chong, S. H. Kim, and S. Y. Lee, Appl. Phys. Lett. **97**, 252112 (2010).

34) J. Y. Choi, K. Heo, K. S. Cho, S. W. Hwang, J. Chung, S. Kim, B. H. Lee, and S. Y. Lee, Sci Rep **7**, 15392 (2017).

35) E. Chong, Y. S. Chun, and S. Y. Lee, Appl. Phys. Lett. **97**, 102102 (2010).

36) S. Tomai, M. Nishimura, M. Itose, M. Matuura, M. Kasami, S. Matsuzaki, H. Kawashima, F. Utsuno, and K. Yano, Jpn. J. Appl. Phys. **51**, 03CB01 (2012).

37) H.-W. Park, B.-K. Kim, J.-S. Park, and K.-B. Chung, Appl. Phys. Lett. **102**, 102102 (2013).

38) N. Xiong, P. Xiao, M. Li, H. Xu, R. Yao, S. Wen, and J. Peng, Appl. Phys. Lett. **102**, 242102 (2013).

39) T. Kizu, N. Mitoma, M. Miyanaga, H. Awata, T. Nabatame, and K. Tsukagoshi, J. Appl. Phys. **118**, 125702 (2015).

40) J.-S. Park, K. Kim, Y.-G. Park, Y.-G. Mo, H. D. Kim, and J. K. Jeong, Adv. Mater. **21**, 329 (2009).

41) S. Parthiban and J.-Y. Kwon, J. Mater. Res. **29**, 1585 (2014).

42) E. Fortunato, P. Barquinha, and R. Martins, Adv. Mater. **24**, 2945 (2012).

43) J. S. Park, W.-J. Maeng, H.-S. Kim, and J.-S. Park, Thin Solid Films **520**, 1679 (2012).

44) H. Jeon, S. Na, M. R. Moon, D. Jung, H. Kim, and H.-J. Lee, J. Electrochem. Soc. **158**, H949 (2011).

45) T. Iwasaki, N. Itagaki, T. Den, H. Kumomi, K. Nomura, T. Kamiya, and H. Hosono, Appl. Phys. Lett. **90**, 242114 (2007).

46) J.-Y. Huh, J.-H. Jeon, H.-H. Choe, K.-W. Lee, J.-H. Seo, M.-K. Ryu, S.-H. K. Park, C.-S. Hwang, and W.-S. Cheong, Thin Solid Films **519**, 6868 (2011).

47) H.-K. Noh, K. Chang, B. Ryu, and W.-J. Lee, Phys. Rev. B **84**, 115205 (2011).

48) K.-H. Liu, T.-C. Chang, K.-C. Chang, T.-M. Tsai, T.-Y. Hsieh, M.-C. Chen, B.-L. Yeh, and W.-C. Chou, Appl. Phys. Lett. **104**, 103501 (2014).

49) X. Gao, M.-F. Lin, B.-H. Mao, M. Shimizu, N. Mitoma, T. Kizu, W. Ou-Yang, T. Nabatame, Z. Liu, and K. Tsukagoshi, J. Phys. D: Appl. Phys. **50**, 025102 (2016).

50) D. Kang, H. Lim, C. Kim, I. Song, J. Park, Y. Park, and J. Chung, Appl. Phys. Lett. **90**, 192101 (2007).

51) W.-T. Chen, S.-Y. Lo, S.-C. Kao, H.-W. Zan, C.-C. Tsai, J.-H. Lin, C.-H. Fang, and C.-C. Lee, IEEE Electron Device Lett. **32**, 1552 (2011).

52) C.-S. Fuh, P.-T. Liu, Y.-T. Chou, L.-F. Teng, and S. M. Sze, ECS J. Solid State Sci. Technol. **2**, Q1 (2012).

53) S.-Y. Sung, J. H. Choi, U. B. Han, K. C. Lee, J.-H. Lee, J.-J. Kim, W. Lim, S. J. Pearton, D. P. Norton, and Y.-W. Heo, Appl. Phys. Lett. **96**, 102107 (2010).

54) K.-S. Son, T.-S. Kim, J.-S. Jung, M.-K. Ryu, K.-B. Park, B.-W. Yoo, K. Park, J.-Y. Kwon, S.-Y. Lee, and J.-M. Kim, Electrochem. Solid-State Lett. **12**, H26 (2009).

55) J. Park, S. Kim, C. Kim, S. Kim, I. Song, H. Yin, K.-K. Kim, S. Lee, K. Hong, J. Lee, J. Jung, E. Lee, K.-W. Kwon, and Y. Park, Appl. Phys. Lett. **93**, 053505 (2008).

56) X. Gao, S. Aikawa, N. Mitoma, M.-F. Lin, T. Kizu, T. Nabatame, and K. Tsukagoshi, Appl. Phys. Lett. **105**, 023503 (2014).

57) N. Mitoma, S. Aikawa, W. Ou-Yang, X. Gao, T. Kizu, M.-F. Lin, A. Fujiwara, T. Nabatame, and K. Tsukagoshi, Appl. Phys. Lett. **106**, 042106 (2015).

58) S. Astha, R. Balasubramaniam, and A. Paranipe, J. Mater. Sci. Lett. **18**, 1555 (1999).

59) Y. Abe, N. Ishiyama, H. Kuno, and K. Adachi, J. Mater. Sci. **40**, 1611 (2005).

60) Y. Abe and N. Ishiyama, Mater. Lett. **61**, 566 (2007).

61) L. T. Yan and R. E. I. Schropp, Thin Solid Films **520**, 2096 (2012).

62) R. K. Gupta, K. Ghosh, and P. K. Kahol, Appl. Surf. Sci. **255**, 8926 (2009).

63) K. Tsukagoshi, J. Tanabe, I. Yagi, K. Shigeto, K. Yanagisawa, and Y. Aoyagi, J. Appl. Phys. **99**, 064506 (2006).

64) Z. Hu, J. Zhang, X. Chen, S. Ren, Z. Hao, X. Geng, and Y. Zhao, Sol. Energy Mater. Sol.

Cells **95**, 2173 (2011).

65)　T. Takenobu, T. Takahashi, T. Kanbara, K. Tsukagoshi, Y. Aoyagi, and Y. Iwasa, Appl. Phys. Lett. **88**, 033511 (2006).

66)　P. T. Liu, C. H. Chang, and C. J. Chang, ECS Trans. **67**, 9 (2015).

67)　Q. Zhang, Z. Yang, M. Qu, R. Fu, P.-T. Liu, and H.-P. D. Shieh, SID Symp. Dig. Tech. Pap. **49**, 225 (2018).

68)　P. Kuo, C. Chang, and P. Liu: 2018 IEEE Symposium on VLSI Technology, 2018,　p. 21.

69)　D.-B. Ruan, P.-T. Liu, K.-J. Gan, Y.-C. Chiu, M.-C. Yu, T.-C. Chien, Y.-H. Chen, P.-Y. Kuo, and S. M. Sze, Thin Solid Films **666**, 94 (2018).

70)　P.-T. Liu, Y.-T. Chou, L.-F. Teng, F.-H. Li, and H.-P. Shieh, Appl. Phys. Lett. **98**,　(2011).

71)　C.-Y. Wu, H.-C. Cheng, C.-L. Wang, T.-C. Liao, P.-C. Chiu, C.-H. Tsai, C.-H. Fang, and C.-C. Lee, Appl. Phys. Lett. **100**, 152108 (2012).

72)　R. D. Shannon, Acta Crystallogr. **32**, 751 (1976).

73)　H. Yabuta, M. Sano, K. Abe, T. Aiba, T. Den, H. Kumomi, K. Nomura, T. Kamiya, and H. Hosono, Appl. Phys. Lett. **89**, 112123 (2006).

74)　H. Kumomi, K. Nomura, T. Kamiya, and H. Hosono, Thin Solid Films **516**, 1516 (2008).

75)　P. F. Carcia, R. S. McLean, M. H. Reilly, and G. Nunes, Appl. Phys. Lett. **82**, 1117 (2003).

76)　H. Hosono, J. Non-Cryst. Solids **352**, 851 (2006).

77)　Y.-R. Luo: *Comprehensive Handbook of Chemical Bond Energies* (CRC Press, 2007) Chap. 14, p. 667.

78)　Y.-R. Luo: *Comprehensive Handbook of Chemical Bond Energies* (CRC Press, 2007) Chap. 16, p. 713.

79)　Y.-R. Luo: *Comprehensive Handbook of Chemical Bond Energies* (CRC Press, 2007) Chap. 9, p. 455.

80)　Y. Q. Jia, J. Solid State Chem. **95**, 184 (1991).

81)　Y. Kang, S. Lee, H. Sim, C. H. Sohn, W. G. Park, S. J. Song, U. K. Kim, C. S. Hwang, S. Han, and D.-Y. Cho, J. Mater. Chem. C **2**, 9196 (2014).

82)　A. Kotani and Y. Toyozawa, J. Phys. Soc. Jpn. **37**, 912 (1974).

83)　M. Campagna, G. K. Wertheim, H. R. Shanks, F. Zumsteg, and E. Banks, Phys. Rev. Lett. **34**, 738 (1975).

84)　E. Chong, Y. W. Jeon, Y. S. Chun, D. H. Kim, and S. Y. Lee, Thin Solid Films **519**, 4347 (2011).

85)　Y.-R. Luo: *Comprehensive Handbook of Chemical Bond Energies* (CRC Press, 2007) Chap. 23, p. 1041.

86)　I. Kang, C. H. Park, E. Chong, and S. Y. Lee, Curr. Appl. Phys. **12**, S12 (2012).

87)　N. Mitoma, B. Da, H. Yoshikawa, T. Nabatame, M. Takahashi, K. Ito, T. Kizu, A. Fujiwara, and K. Tsukagoshi, Appl. Phys. Lett. **109**, 221903 (2016).

88)　B. Kim, E. Chong, D. Hyung Kim, Y. Woo Jeon, D. Hwan Kim, and S. Yeol Lee, Appl. Phys. Lett. **99**, 062108 (2011).

89)　D. Hyung Kim, D. Youn Yoo, H. Kwang Jung, D. Hwan Kim, and S. Yeol Lee, Appl. Phys. Lett. **99**, 172106 (2011).

90)　Y. S. Jung, J. Y. Seo, D. W. Lee, and D. Y. Jeon, Thin Solid Films **445**, 63 (2003).

91)　H. Q. Chiang, B. R. McFarlane, D. Hong, R. E. Presley, and J. F. Wager, J. Non-Cryst. Solids **354**, 2826 (2008).

92)　K. H. Ji, J.-I. Kim, H. Y. Jung, S. Y. Park, R. Choi, U. K. Kim, C. S. Hwang, D. Lee, H. Hwang, and J. K. Jeong, Appl. Phys. Lett. **98**, 103509 (2011).

93)　K. Ghaffarzadeh, A. Nathan, J. Robertson, S. Kim, S. Jeon, C. Kim, U. I. Chung, and J.-H. Lee, Appl. Phys. Lett. **97**, 113504 (2010).

94)　R. Cross and M. De Souza, Appl. Phys. Lett. **89**, 263513 (2006).

95)　K. Hoshino, D. Hong, H. Q. Chiang, and J. F. Wager, IEEE Trans. Electron Devices **56**, 1365 (2009).

96)　S. Urakawa, S. Tomai, Y. Ueoka, H. Yamazaki, M. Kasami, K. Yano, D. Wang, M. Furuta, M. Horita, Y. Ishikawa, and Y. Uraoka, Appl. Phys. Lett. **102**, 053506 (2013).

97)　M. Fujii, Y. Uraoka, T. Fuyuki, J. S. Jung, and J. Y. Kwon, Jpn. J. Appl. Phys. **48**, 04C091

(2009).

98) R. B. M. Cross, M. M. De Souza, S. C. Deane, and N. D. Young, IEEE Trans. Electron Devices **55**, 1109 (2008).

99) K. Takechi, M. Nakata, T. Eguchi, H. Yamaguchi, and S. Kaneko, Jpn. J. Appl. Phys. **48**, 078001 (2009).

100) W. v. Meyer and H. Neldel, Z. tech. Phys **18**, 588 (1937).

101) J. Jeong, J. K. Jeong, J.-S. Park, Y.-G. Mo, and Y. Hong, Jpn. J. Appl. Phys. **49**, 03CB02 (2010).

102) S. Lee and A. Nathan, Appl. Phys. Lett. **101**, 113502 (2012).

103) C. Chen, K. Abe, H. Kumomi, and J. Kanicki, IEEE Trans. Electron Devices **56**, 1177 (2009).

104) J. Stuke, J. Non-Cryst. Solids **97**, 1 (1987).

105) W. Jackson, Phys. Rev. B **38**, 3595 (1988).

106) E. Meijer, M. Matters, P. Herwig, D. De Leeuw, and T. Klapwijk, Appl. Phys. Lett. **76**, 3433 (2000).

107) L.-F. Mao, H. Ning, C. Hu, Z. Lu, and G. Wang, Sci. Rep. **6**, 24777 (2016).

108) R. Flückiger, J. Meier, M. Goetz, and A. Shah, J. Appl. Phys. **77**, 712 (1995).

109) M. Kimura, T. Nakanishi, K. Nomura, T. Kamiya, and H. Hosono, Appl. Phys. Lett. **92**, (2008).

110) K. Ide, Y. Kikuchi, K. Nomura, M. Kimura, T. Kamiya, and H. Hosono, Appl. Phys. Lett. **99**, 093507 (2011).

111) K. Nomura, T. Kamiya, and H. Hosono, ECS J. Solid State Sci. Technol. **2**, P5 (2012).

112) S. D. Wang, T. Minari, T. Miyadera, K. Tsukagoshi, and Y. Aoyagi, Appl. Phys. Lett. **91**, 203508 (2007).

113) J. Park, C. Kim, S. Kim, I. Song, S. Kim, D. Kang, H. Lim, H. Yin, R. Jung, E. Lee, J. Lee, K.-W. Kwon, and Y. Park, IEEE Electron Device Lett. **29**, 879 (2008).

114) W.-F. Chung, T.-C. Chang, C.-S. Lin, K.-J. Tu, H.-W. Li, T.-Y. Tseng, Y.-C. Chen, and Y.-H. Tai, J. Electrochem. Soc. **159**, H286 (2012).

115) Y. Ueoka, Y. Ishikawa, J. P. Bermundo, H. Yamazaki, S. Urakawa, Y. Osada, M. Horita, and Y. Uraoka, Jpn. J. Appl. Phys. **53**, 03CC04 (2014).

116) K.-H. Choi and H.-K. Kim, Appl. Phys. Lett. **102**, 052103 (2013).

117) T. Minari, T. Miyadera, K. Tsukagoshi, Y. Aoyagi, and H. Ito, Appl. Phys. Lett. **91**, 053508 (2007).

118) T. Matsumoto, W. Ou-Yang, K. Miyake, T. Uemura, and J. Takeya, Org. Electron. **14**, 2590 (2013).

119) S. Aikawa, N. Mitoma, T. Kizu, T. Nabatame, and K. Tsukagoshi, Appl. Phys. Lett. **106**, 192103 (2015).

120) S. Kim, K.-K. Kim, and H. Kim, Appl. Phys. Lett. **101**, 033506 (2012).

121) M. Kimura, T. Kamiya, T. Nakanishi, K. Nomura, and H. Hosono, Appl. Phys. Lett. **96**, 262105 (2010).

122) T. Kamiya, K. Nomura, and H. Hosono, Appl. Phys. Lett. **96**, 122103 (2010).

123) T. Kamiya, K. Nomura, and H. Hosono, J. Disp. Technol. **5**, 462 (2009).

124) S. Lee, K. Ghaffarzadeh, A. Nathan, J. Robertson, S. Jeon, C. Kim, I. H. Song, and U. I. Chung, Appl. Phys. Lett. **98**, 203508 (2011).

125) S. Sallis, K. T. Butler, N. F. Quackenbush, D. S. Williams, M. Junda, D. A. Fischer, J. C. Woicik, N. J. Podraza, B. E. White, A. Walsh, and L. F. J. Piper, Appl. Phys. Lett. **104**, 232108 (2014).

126) J. K. Jeong, H. Won Yang, J. H. Jeong, Y.-G. Mo, and H. D. Kim, Appl. Phys. Lett. **93**, 123508 (2008).

127) P.-T. Liu, Y.-T. Chou, and L.-F. Teng, Appl. Phys. Lett. **95**, 233504 (2009).

128) W.-F. Chung, T.-C. Chang, H.-W. Li, S.-C. Chen, Y.-C. Chen, T.-Y. Tseng, and Y.-H. Tai, Appl. Phys. Lett. **98**, 152109 (2011).

129) E. D. Specht, A. Goyal, and D. M. Kroeger, Supercond. Sci. Technol. **13**, 592 (2000).

130) X. Guo, S. R. P. Silva, and T. Ishii, Appl. Phys. Lett. **93**, 042105 (2008).

131) H. Tang, K. Ishikawa, K. Ide, H. Hiramatsu, S. Ueda, N. Ohashi, H. Kumomi, H. Hosono, and T. Kamiya, J. Appl. Phys. **118**, 205703 (2015).

132) T. Kizu, S. Aikawa, T. Nabatame, A. Fujiwara, K. Ito, M. Takahashi, and K. Tsukagoshi, J. Appl. Phys. **120**, 045702 (2016).

133) N. Birks, G. H. Meier, and F. S. Pettit: *Introduction to the high temperature oxidation of metals* (Cambridge University Press, 2006).

134) K. Nomura, T. Kamiya, and H. Hosono, Thin Solid Films **520**, 3778 (2012).

135) S. T. Kim, Y. Shin, P. S. Yun, J. U. Bae, I. J. Chung, and J. K. Jeong, Electron. Mater. Lett. **13**, 406 (2017).

136) K. Park, H.-W. Park, H. S. Shin, J. Bae, K.-S. Park, I. Kang, K.-B. Chung, and J.-Y. Kwon, IEEE Trans. Electron Devices **62**, 2900 (2015).

137) B. Ryu, H.-K. Noh, E.-A. Choi, and K. J. Chang, Appl. Phys. Lett. **97**, 022108 (2010).